\documentclass[twocolumn,pre,floatfix]{revtex4}

\usepackage{amsmath}
\usepackage{amssymb}
\usepackage{bm}                          
\usepackage{ifthen}
\usepackage{color}
\usepackage{graphicx}

\DeclareMathAlphabet{\matheurm}{U}{eur}{m}{n}

\newcommand\eq[1]                              
{
\begin{align*}
#1
\end{align*}
}

\newcommand\eql[2] 
{
\begin{equation}\label{#1}
\begin{split}
#2
\end{split}
\end{equation}
}

\newcommand\eqnl[2]        
{
\begin{equation}\label{#1}
#2
\end{equation}
}

\newcommand\eqsl[2]                            
{
\begin{align}\label{#1}
#2
\end{align}
}

\newcommand\eqssl[2]                      
{
\begin{subequations}\label{#1}
\begin{align}
#2
\end{align}
\end{subequations}
}

\renewcommand\_[1]     {_{#1}^{}}
\newcommand\vgrk[1]    {\pmb{\bm{#1}}}   
\newcommand\DRAFTWORK[1] {}

\providecommand\eqref[1] {(\ref{#1})}  
\newcommand\Eq[1]      {Eq.~\eqref{#1}}
\newcommand\Eqs[1]     {Eqs.~\eqref{#1}}
\newcommand\Fig[1]     {Fig.~\ref{#1}}
\newcommand\Sec[1]     {Sec.~\ref{#1}}
\newcommand\Ref[1]     {Ref.~\onlinecite{#1}}
\newcommand\Refs[1]    {Refs.~\onlinecite{#1}}
\newcommand\Cite[1]    {~\cite{#1}}                

\newcommand\ME[3]      {\langle{{#1}}|{{#2}}|{{#3}}\rangle} 
\newcommand\ket[1]     {|{{#1}}\rangle}
\newcommand\bra[1]     {\langle{{#1}}|}
\newcommand\braket[2]  {\langle{{#1}}|{{#2}}\rangle}

\newcommand\Matrix[1]                    
{{\begin{pmatrix} #1 \end{pmatrix}}}

\newcommand\matrixr[2][]
                       {{\bm{\mathsf{#2}}#1}}

\newcommand\matrixg[2][]
                       {{\bm{#2}#1}}             
\newcommand\matrixgT[2][]
                       {{\bm{#2}#1^{\mathrm{T}}}}             
\newcommand\matrixgHC[2][]                           
                       {{\bm{#2}#1^\dagger}}

\newcommand\per[1]     {\mathrm{per}\left({#1}\right)}

\newcommand\Db[1]      {b_{{#1}}^{}}
\newcommand\Cb[1]      {b_{{#1}}^\dagger}
\newcommand\Dc[1]      {c_{{#1}}^{}}
\newcommand\Cc[1]      {c_{{#1}}^\dagger}

\newcommand\Cphi[1]    {\hat{\phi}_{{#1}}^\dagger}
\newcommand\Dpsi[1]    {\hat{\psi}_{{#1}}^{}}
\newcommand\Cpsi[1]    {\hat{\psi}_{{#1}}^\dagger}

\newcommand\Pgs        {\mathcal{P}_\mathrm{gs}}      
\newcommand\zPgs       {\widetilde{\mathcal{P}}_\mathrm{gs}}
\newcommand\Aop        {{\hat{A}}}
\newcommand\Bop        {{\hat{B}}}
\newcommand\BVop       {{\hat{B}_v}}
\newcommand\HSop       {{\hat{v}}}
\newcommand\Hop        {{\hat{H}}}
\newcommand\Kop        {{\hat{K}}}
\newcommand\Top        {{\hat{T}}}
\newcommand\Vop        {{\hat{V}}}

\newcommand\nop[1]     {{\hat{n}\_{#1}}}
\newcommand\nbar       {\bar{n}} 

\newcommand\XV[1]      {\langle{#1}\rangle}

\newcommand\kinXV      {\XV{\Top}}
\newcommand\trapXV     {\XV{\Vop_{\mathrm{trap}}}}
\newcommand\potXV      {\XV{\Vop_{\mathrm{2B}}}}

\newcommand\trialEst[1]{{#1}_{\mathrm{T}}}
\newcommand\mixedEst[1]{{#1}_{\mathrm{mix}}}
\newcommand\bruteEst[1]{{#1}_{\mathrm{bf}}}
\newcommand\BPEst[1]   {{#1}_{\mathrm{bp}}}
\newcommand\trialXV[1] {\trialEst{\XV{#1}}}
\newcommand\mixedXV[1] {\mixedEst{\XV{#1}}}
\newcommand\bruteXV[1] {\bruteEst{\XV{#1}}}
\newcommand\BPXV[1]    {\BPEst{\XV{#1}}}

\newcommand\Vphi[2][]  {{\vgrk{\phi}_{#2}#1^{}}}  
\newcommand\Vpsi[2][]  {{\vgrk{\psi}_{#2}#1^{}}}

\newcommand\VpsiT[2][] {{\vgrk{\psi}_{#2}#1^{\dagger}}}        

\newcommand\half       {{\textstyle\frac{1}{2}}}
\newcommand\Half       {\frac{1}{2}}

\newcommand\intdr[1][] {\! \int \! d^3\rvec#1 \,}

\newcommand\expP[1]    {\exp \! \left( {{#1}} \right)} 
\newcommand\expB[1]    {\exp \! \left[ {{#1}} \right]} 

\newcommand\aho[1][]   {a_{\mathrm{ho}}^{#1}}            
\newcommand\as[1][]    {a_s^{#1}}              
\newcommand\ET         {E_\mathrm{T}}                       
\newcommand\EL         {E_\mathrm{L}^{}}         
\newcommand\PsiT[1][]  {\Psi_{\mathrm{T}#1}^{}}
\newcommand\psiT[1][]  {\psi_{\mathrm{T}#1}^{}}
\newcommand\PsiGP[1][] {\Phi_{\mathrm{GP}}^{#1}}
\newcommand\PsiGS      {\Phi_0}
\newcommand\PsiGSBP    {\Phi_0^{\mathrm{bp}}}

\newcommand\Dt         {\Delta\tau}                  
\newcommand\tauBP      {\tau_{\mathrm{bp}}^{}}      
\newcommand\nBP        {n_{\mathrm{bp}}^{}}          
\newcommand\expdtK     {\exp \! {\big( \! - \! \half \Dt \Kop \big)}}

\newcommand\expdtV     {\exp \! {\big( \! - \! \Dt \Vop \big)}}
\newcommand\expdtH     {\exp \! {\big( \! - \! \Dt \Hop \big)}}
\newcommand\exptBPH    {\exp \! {\big( \! - \! \tau_{\mathrm{bp}}^{}
                                       \Hop \big)}}
\newcommand\Order      {\mathcal{O}}
\newcommand\spin[1]    {#1}       
\newcommand\spinsum[1] {\sum_{#1}}                             

\newcommand\Tcm        {T_{\mathrm{cm}}}
\newcommand\tcm        {t_{\mathrm{cm}}}
\newcommand\Topcm      {{\hat{T}_{\mathrm{cm}}}}        
\newcommand\PsiCM      {\Psi_{\mathrm{cm}}}         

\newcommand\xbar[1][]  {\underline{x}_{#1}}
\newcommand\Vxbar[1][] {\vec{\underline{x}}_{#1}}
\newcommand\xsqrbar[1][] 
                       {\underline{{x}_{#1}^2}}
\newcommand\vbar[1][]  {\bar{v}_{#1}}
\newcommand\vsqrbar[1][] {\overline{{v}_{#1}^2}}

\newcommand\wlkridx[1] {\matheurm{#1}}               

\newcommand\wIDX[3][]  {
                        \ifthenelse{\equal{#3}{}}
                                   {_{\wlkridx{#2}}#1}
                                   {_{\wlkridx{#2}}^{(#3)}{}#1}
                       }
\newcommand\wlkr[3][]  {\phi\wIDX[#1]{#2}{#3}}

\newcommand\Wlkr[3][]  {\ket{\wlkr[#1]{#2}{#3}}}

\newcommand\Ovlp[3][]  {w\wIDX[#1]{#2}{#3}}
\newcommand\zOvlp[3][] {{w}\wIDX[#1]{#2}{#3}}

\newcommand\bpath[2]   {\Bop_{\wlkridx{#1}}^{(#2)}}
\newcommand\wpath[2]   {W_{\wlkridx{#1}}^{(#2)}}

\newcommand\bwlkr[3][] {
                        \eta_{\wlkridx{#2}}
                        ^{\ifthenelse{\equal{#3}{}}{}{(#3)}}
                        {}#1
                       }
\newcommand\bWlkr[3][] {\ket{\bwlkr[#1]{#2}{#3}}}
\newcommand\bOvlp[3][] {
                        u_{\wlkridx{#2}}
                        ^{\ifthenelse{\equal{#3}{}}{}{(#3)}}
                        {}#1
                       }

\newcommand\Nwlkr      {N_{\mathrm{wlkr}}}

\newcommand\rvec       {\mathbf{r}}                           

\newcommand\fmtSC[1]   {}
\newcommand\fmtTC[1]   {#1}

\fmtTC{
\renewcommand\expP[1]    {e^{
    \renewcommand\half {\frac{1}{2}}
    #1
}} 
\renewcommand\expB[1]    {e^{
    \renewcommand\half {\frac{1}{2}}
    #1
}} 
\renewcommand\expdtK     {e^{-\Half \Dt \Kop}}

\renewcommand\expdtV     {e^{-\Dt \Vop}}
\renewcommand\expdtH     {e^{-\Dt \Hop}}
\renewcommand\exptBPH    {e^{-\tau_{\mathrm{bp}}^{} \Hop}}
} 

\renewcommand\spin[1]    {}
\renewcommand\spinsum[1] {}

\definecolor{Green}{rgb}{0.2,0.96,0.2}
\definecolor{Remarks}{rgb}{1,0.3,0.3}
\definecolor{Extra}{rgb}{0.2,0.2,1}
\definecolor{Blue}{rgb}{0.2,0.3,1}
\definecolor{Black}{rgb}{0,0,0}

\newcommand\COMMENTED[1] {}
\newcommand\EXTRA[2]     {\color{Extra}#1\color{Black}} 

\newcommand\PLOTCOLOR[2] {#2}            
\newcommand\PLOTFILE[1]  {#1}

\renewcommand\EXTRA[2]   {#2}

\begin{document} 

\title{Quantum Monte Carlo method for the ground state of many-boson
systems}
\date{\today}

\author{Wirawan Purwanto} \email{wirawan@camelot.physics.wm.edu}
\author{Shiwei Zhang}     \email{shiwei@physics.wm.edu}
\affiliation{Department of Physics,
             The College of William and Mary,
             Williamsburg, Virginia 23187}

\begin{abstract}

We formulate a quantum Monte Carlo (QMC) method for calculating the ground state
of many-boson systems. The method is based on a field-theoretical
approach, and is closely related to existing fermion auxiliary-field QMC
methods which are applied in several fields of physics. The ground-state
projection is implemented as a branching random walk in the space of
permanents consisting of identical single-particle orbitals.
Any single-particle basis can be used, and the method is in principle exact.
We illustrate this method with a trapped atomic boson gas, where the atoms
interact via an attractive or repulsive contact two-body potential.
We choose as the single-particle basis a real-space grid.
We compare with
exact results in small systems, and arbitrarily-sized systems of
untrapped bosons with attractive interactions in one dimension, where
analytical solutions exist.
We also compare with the corresponding Gross-Pitaevskii (GP) mean-field
calculations for trapped atoms, and discuss the close formal relation
between our method and the GP approach. Our method provides a way to
systematically improve upon GP while using the same framework, capturing
interaction and correlation effects with a stochastic, coherent ensemble
of non-interacting solutions.
We discuss various algorithmic issues, including importance sampling and
the back-propagation technique for computing observables, and illustrate
them with numerical studies.
We show results for systems with up to $N\sim 400$ bosons.

\end{abstract}

\COMMENTED{
\begin{verbatim}
TBHQMC PAPER DRAFT --------------------------------------------
CVS $Id: tbh-paper.tex,v 1.51.2.2 2004/03/31 01:37:25 wirawan Exp $
---------------------------------------------------------------



\end{verbatim}
}

\maketitle 

\section{Introduction}\label{sec:Introduction} 

The study of many-body quantum systems has been a very challenging
research field for many years. Computational methods have often been the
way of choice to extract theoretical understanding on such systems.
Most computational quantum mechanical studies are based on simpler
mean-field theories such as the Gross-Pitaevskii (GP) equation for
bosons or the Kohn-Sham density-functional theory (DFT) for fermions.
Despite their remarkable success, the treatment of particle interaction or
correlation effects is only approximate within these approaches, and can
lead to incorrect results, especially as the strength of particle
interactions is increased. It is therefore necessary to develop
alternative computational methods that can describe the effect of
interaction more accurately and reliably.

In this paper we present a quantum Monte Carlo (QMC) method to study the
ground state of many-boson systems. The method is in principle exact. Our
interest in the development and use of this method was motivated by the
realization of the Bose-Einstein condensation in ultracold atomic
gases\Cite{Anderson1995}. These are dilute gases consisting of interacting
alkali atoms. The interaction among the atoms is well described by a
simple two-body potential, either attractive or repulsive, based on the
scattering length. For weakly-interacting systems the mean-field GP
approach has, as expected, performed extremely
well\Cite{Dalfovo99,Leggett01}.
More recently, Fesbach resonances\Cite{Cornish2000} have successfully been
used as a powerful way to tune the strength of the interaction
experimentally. This provides a source of rich physics, and increases the
need for theoretical methods which can benchmark GP and provide an
alternative where GP is inadequate.

Several QMC methods exist for calculating the properties of interacting
many-body systems. The ground-state diffusion Monte Carlo\Cite{Foulkes2001}
and the finite-temperature path-integral Monte Carlo
(PIMC)\Cite{Ceperley1995} methods, which work in
many-particle configuration space and in the first-quantized framework,
have been successfully applied to a variety of boson and fermion
systems.
In the context of atomic gases, Krauth\Cite{Krauth1996},
Gruter \textit{et al.}\Cite{Gruter1997}, and Holzmann and
Krauth\Cite{Holzmann1999} have employed PIMC to study finite-temperature
properties of trapped bosons with positive scattering lengths, modeling the
two-body interactions by a hard-sphere potential.
Glyde and co-workers have studied the ground state of trapped bosons,
also by hard spheres\Cite{DuBois2001,DuBois2003}.
Ulmke and Scalletar\Cite{Ulmke2000} did finite-temperature QMC
calculations on quantum spin systems and the Bose-Hubbard model. In the
latter calculation, a hard-core repulsive potential was assumed, which
allowed a transformation of the problem into an XXZ spin-like
problem that can be treated with a fermion QMC method.

Our method is based on the auxiliary field quantum Monte Carlo (AFQMC)
approach\Cite{Blankenbecler1981,Koonin1986}.
The AFQMC is a field-theoretical method, where many-body propagators
resulting from two-body interactions are transformed, by use of auxiliary
fields, into a many-dimensional integral over one-body propagators
\Cite{Hubbard1959,Stratonovich1957}. The many-dimensional integral is then
computed using stochastic means.\EXTRA{\ The many-body wave function is
expressed in terms of a single-particle basis, usually chosen to maximize
the sampling efficiency.~}{}
The AFQMC framework is appealing for several reasons. Working in
second-quantization, it automatically imposes the proper
particle-permutation symmetry or antisymmetry.
It provides a many-body method with close formal relation to mean-field
approaches, as we discuss later. In addition, it allows convenient
calculation of the observables and correlation functions.

The AFQMC method has been widely employed to study fermion systems in
condensed matter\Cite{Hirsch1983,White1989,Zhang1997_CPMC}, nuclear
physics\Cite{Dean1999,Fantoni2001}, and lattice gauge theory.
In this paper, we generalize the fermion ground-state
auxiliary-field quantum Monte Carlo method\Cite{Zhang1997_CPMC,Zhang2003}
to many-boson systems.
We project the many-body boson ground-state from an initial trial state
$\ket{\PsiT}$. Our choice of $\ket{\PsiT}$ is a permanent consisting of
$N$ identical single-particle orbitals, which was first suggested in a
model calculation by Sugiyama and Koonin\Cite{Koonin1986}.
The many-body ground state is projected from $\ket{\PsiT}$ with
open-ended, branching random walks to sample the auxiliary fields. We
formulate an importance sampling scheme, which greatly improves the
efficiency of the method and makes possible simulations of large systems.
We also discuss in detail the back-propagation technique which allows
convenient calculation of virtually any ground-state observables.

Our method retains all the advantages of AFQMC. It allows the use of any
single-particle basis, which in this paper is chosen to be a real-space
grid.
As we discuss in Sec.~\ref{sec:Discussions}, it
provides a means for true many-body calculations in a framework which
closely relates to the GP approach. The approach can be viewed as a
stochastic collection of parallel GP-like calculations whose ``coherent''
linear combination gives the interaction and correlation effects.

In this paper we present our QMC method for bosons and discuss its
behavior and characteristics.
We use a trapped atomic boson gas as our test system, where the atoms
interact via an attractive or repulsive contact two-body potential.
A sufficiently detailed description of the method is given to facilitate
implementation.
Compared to its fermionic counterpart, our method here is
formally simpler.
It therefore also offers opportunities to study algorithmic
issues. Because of the intense interest in methods for treating correlated
systems (fermions or bosons) and the relatively early stage of this
type of QMC methods, a second purpose of the paper is to use the bosonic
test ground to explore, discuss, and illustrate the generic features of
ground-state QMC methods based on auxiliary fields.
An example is the case of repulsive interactions, where a
phase problem appears in a bosonic system, which provides a clean
test ground to study methods for controlling this problem\Cite{Zhang2003},
which is crucial for applications in fermion systems.
The majority of the applications in this paper will be to systems where
exact results are available for benchmark. These include small systems,
which can be diagonalized exactly, and the case of untrapped bosons with
attractive interactions in one dimension, where analytical solutions
exist. It is worth emphasizing that the method scales gracefully (similar
to GP) and allows calculations for a large number ($N$) of bosons. We will
show results for larger systems ($\sim 1000$ sites and hundreds of
particles) in one- and three-dimensions to illustrate this.

Our paper is organized as follows.
In section \ref{sec:Background}, we establish some conventions and review
the basic ground-state projection and auxiliary-field quantum Monte Carlo
method.
In section \ref{sec:rnd-wlk}, we introduce our new AFQMC implementation
for bosons, including the formulation of an importance-sampling scheme and
the back-propagation technique for convenient calculation of virtually any
ground-state observables.
In section \ref{sec:TBH}, we describe the implementation of our method to
study the ground state of a trapped Bose atomic gas, which we model by by
a Bose-Hubbard Hamiltonian with an external trapping potential. We also
describe our implementation of the GP approach to study the same
Hamiltonian.
In section \ref{sec:DemoResults}, we present our computational results. We
benchmark the method in systems where exact results are available. We also
provide examples to illustrate the behavior and key characteristics of our
method. We carry out GP calculations on the same Hamiltonian and compare
the results with those from our QMC calculations.
In section \ref{sec:Discussions} we comment on some characteristics of the
method, further discuss its relation to and differences from GP, and
mention future directions and some immediate applications of this method.
Some computing issues will also be discussed.
Finally, in the appendices we provide additional technical details of the
method.

\section{Background}\label{sec:Background} 

\subsection{Many-body Hamiltonian}

We use the second quantized formalism throughout this paper. We assume
that an appropriate set of single-particle basis
$\{\ket{\chi\_i}\}$ has been chosen, in terms of which the wave functions
will be expanded. For simplicity, we assume that the single-particle
basis is orthonormal, although this is not required.
The number of basis states is $M$.
The operators $\Cc{i}$ and $\Dc{i}$,
respectively, are the usual creation and annihilation operator for the
state $\ket{\chi\_i}$. They satisfy the commutation relation
$
    [\Dc{i}, \Cc{j}]\_- = \delta\_{ij}
$.
This automatically imposes the symmetrization requirement of the
many-body wave functions.

We limit our discussion to a quantum-mechanical, many-body
system with two-body interactions.
The Hamiltonian $\Hop$ has a general form of
\eql{eq:GenericHamiltonian}
{
    \Hop = \Kop + \Vop \,,
}
where $\Kop$ is the sum total of all the one-body operators (the
kinetic energy and external potential energy),
\eq{
    \Kop = \sum_{ij} K\_{ij} \Cc{i}\Dc{j}
\EXTRA{ \,, \\
    K\_{ij} = \ME{\chi\_i}{\Kop}{\chi\_j}
}{} \,;
}
and $\Vop$ contains the two-body interactions:
\eq{
    \Vop = \sum_{ijkl} V\_{ijkl}
           \Cc{i}\Cc{j}\Dc{k}\Dc{l}
\EXTRA{ \,, \\
    V\_{ijkl} = \ME{\chi\_i \chi\_j}{\Vop}{\chi\_l \chi\_k}
}{} \,.
}
Our objective is to calculate the ground state properties of such a
system, which contains a fixed number of particles, $N$.

\subsection{Ground state projection}

The ground state wave function $\ket{\Phi\_0}$ can be readily extracted
from a given trial solution $\ket{\PsiT}$ using the ground-state
projection operator
\eql{eq:def:g.s.proj.op}
{
    \Pgs \equiv \expP{-\Dt\Hop} \expP{\Dt \ET} \,,
}
where $\ET$ is the best guess of the ground-state energy, provided that
$\ket{\PsiT}$ is not orthogonal to $\ket{\Phi\_0}$. Applying the
operator $\Pgs$ repeatedly to the initial wave
function $\ket{\PsiT}$ would exponentially attenuate the excited-state
components of the initial wave function, leaving only the ground state:
\begin{subequations}\label{eq:gsProjection}
\begin{eqnarray}
     (\Pgs)^n_{} \ket{\PsiT}
     &\!\overset{n\rightarrow\infty}\longrightarrow\!& \ket{\Phi_0} \,;\\
     \Pgs \ket{\Phi_0}
     &\longrightarrow& \ket{\Phi_0} \,. \label{eq:gsProjection-2}
\end{eqnarray}
\end{subequations}
Because of its resemblance to the real-time propagator, the operator
$\Pgs$ is also called the imaginary-time propagator.
In ground-state QMC methods, $\Pgs$ is evaluated by means of a Monte Carlo
sampling, resulting in a stochastic representation of the ground-state
wave function.

\subsection{Basic auxiliary-field method}\label{ssec:AFQMC} 

Two essential ingredients are needed in order to evaluate $\Pgs$ within a
reasonable computing time. The first is the
Trotter-Suzuki approximation\Cite{Trotter1959,Suzuki1976}.
The propagator is broken up into a product of exponential operators, which
becomes exact in the limit $\Dt \rightarrow 0$. The second-order form of
this approximation is
\eql{eq:Trotter-Suzuki}
{
    \expB{-\Dt(\Kop + \Vop)}          \fmtTC{&}
  = \,\expdtK{} \expdtV{} \expdtK{}   \fmtTC{\\&}
  + \Order(\Dt^3) \,.
}

The second ingredient is the Hubbard-Stratonovich (HS)
transformation\Cite{Hubbard1959,Stratonovich1957}, which allows us to
reduce the two-body propagator to a multidimensional integral
involving only one-body operators, using the following identity:%
\Cite{Zhang2000_BookChapter}
\eql{eq:HSxformn}
{
    \expP{\half\Dt \HSop^2} =
    \frac{1}{\sqrt{2\pi}} \int_{-\infty}^{\infty} \! dx \,
    \expP{-\half x^2} \expP{x \sqrt{\Dt} \, \HSop} \,,
}
where $\HSop$ is a one-body operator\EXTRA{, which is always quadratic in
the creation and annihilation operator}{}:
\eq{
    \HSop \equiv
    \sum_{ij} \HSop\_{ij} \Cc{i}\Dc{j} \,.
}
%
The hermiticity of $\Vop$ allows us to
decompose it into a sum of the square of one-body operators
$\{\HSop\_i\}$ (see, for example, \Refs{Zhang1997_CPMC} and
\onlinecite{Zhang2000_BookChapter}):
\eql{eq:VopSumSqr}
{
    \Vop = -\half \sum_i \HSop_i^2 \,,
}
Because of this, we can always apply the Hubbard-Stratonovich
transformation on a general two-body potential operator:
\eql{eq:HSxformn-general}
{
    \expP{-\Dt\Vop}
& = \prod_i \expP{\half\Dt \HSop_i^2} + \Order(\Dt^2)
    \\
& = \prod_i \int_{-\infty}^{\infty} \! dx_i
    \frac{\expP{-\half x_i^2}}{\sqrt{2\pi}}
    \expP{x_i \sqrt{\Dt} \, \HSop\_i} + \Order(\Dt^2) \,.
}
In general, the Trotter breakup incurs an additional systematic
error of $\Order(\Dt^2)$.

Applying these two procedures, we obtain an approximate expression of the
ground-state projection operator:
\begin{widetext}
\eql{eq:ProjectionOp}
{
    \Pgs &= \,
    e^{\Dt\ET} \cdot
    \expdtK{}
    \left\{ \prod_{i}
            \int_{-\infty}^{\infty} \! dx\_i \, p(x\_i)
            \expP{x\_i \sqrt{\Dt} \, \HSop\_{i}}
    \right\}
    \expdtK{}
    \fmtSC{\\&}
  +~\Order(\Dt^2)
    \,,
}
\end{widetext}
where $p(x)$ is the normalized Gaussian probability density function
with unit standard deviation:
$
    p(x) \equiv \frac{1}{\sqrt{2\pi}} \expP{-\half x^2}
$.
This approach is applicable to both boson and fermion systems. It enables
us to compute the exact ground state of a quantum many-body system.
To reduce the systematic error from the finite timestep $\Dt$, the
so-called ``Trotter error'', small timesteps $\Dt$ are necessary.
Often, calculations are performed  for several $\Dt$ values, then an
extrapolation to $\Dt \rightarrow 0$ is made to remove the Trotter error.

For convenience we define the following notations:
\begin{itemize}

\item $\vec{x} \equiv \{x\_1, x\_2, \ldots\} \,$: the collection of all
the auxiliary-fields.

\item $p(\vec{x}) \equiv \prod_i p(x\_i) \,$: a (normalized)
multidimensional probability density function\EXTRA{~(MPDF)}{}, which
is the product of the one-dimensional probability density
functions $p(x\_i)$. \EXTRA{In practice, it is also much easier to
sample from such a separable MPDF.}{}

\item $\BVop(\vec{x}) \,$: a product of the exponential one-body operators
arising from the auxiliary-field transformation.
From \Eq{eq:ProjectionOp},
$
    \BVop(\vec{x}) \equiv
    \prod_i \expP{x\_i \sqrt{\Dt}\,\HSop\_i}\,
$.

\item $\Bop(\vec{x})\,$: the product of $\BVop(\vec{x})$ with all other
one-body exponential operators that do not depend on the auxiliary fields
$\vec{x}$, and all the necessary scalar prefactors. For the projector in
Eq.~\eqref{eq:ProjectionOp}, $\Bop(\vec{x}) \equiv e^{\Dt\ET} \cdot
\expdtK\,\BVop(\vec{x})\,\expdtK\,$.

\end{itemize}
With these notations, $\Pgs$ takes a generic form of a high-dimensional
integral operator:
\eql{eq:Pgs-generic}
{
    \Pgs
&   \approx \int d\vec{x} \, p(\vec{x}) \Bop(\vec{x})
    \,.
}
%
%

\subsection{Wave function representation}\label{ssec:wf-reps}


We write our wave functions in terms of the basis functions
$\ket{\chi\_i}$. A single-particle wave function is written as
\eql{eq:one-particle-wf}
{
    \ket{\varphi} = \sum_i \varphi\_i \ket{\chi\_i}
                  = \sum_i \varphi\_i \Cc{i} \ket{0}
             \equiv \hat\varphi^\dagger \ket{0} \,.
}
A single-permanent, $N$-Bosons wave function is given by
\eql{eq:one-permament-many-body-wf}
{
    \ket{\phi} = \Cphi{1} \Cphi{2} \ldots \Cphi{N}\ket{0} \,.
}
In general, the exact ground state wave function is a superposition of
such permanents. Unlike the fermionic case, where the particles occupy
mutually orthogonal orbitals, there is no such restriction on the orbitals
here.
We use this freedom in our method to have all the bosons occupy the same
orbital in $\ket{\phi}$, which greatly simplifies the
computation\Cite{Koonin1986}. We will refer to this as \emph{identical
orbital representation} (IOR).
The most important virtue of this representation is that the exponential
of a one-body operator $\Aop$ transform a single-permanent wave function
$\ket{\phi}$ into another single-permanent wave function
$\ket{\phi'}$:\Cite{Hamann1990}
\eql{eq:exp-operator}
{
    \expP{\Aop} \ket{\phi} = \ket{\phi'} \,.
}
In particular, $\Bop(\vec{x})$ in Eq.~\eqref{eq:exp-operator} transforms a
single permanent $\ket{\phi}$ into another single permanent $\ket{\phi'}$.
(In Appendix \ref{app:IOR} we include a brief summary of properties
of wave functions in IOR.)

\subsection{Metropolis AFQMC}\label{ssec:Metropolis}

Standard AFQMC calculations\Cite{Koonin1986} employ Metropolis Monte Carlo
to compute various ground-state observables,
\begin{widetext}
\eql{eq:MMC-AFQMC}
{
    \XV{\Aop}_{\mathrm{g.s.}}
& = \frac{ \ME{\PsiT}{\Pgs\cdots\Pgs\,\Aop\,\Pgs\cdots\Pgs}{\PsiT} }
         { \ME{\PsiT}{\Pgs\cdots\Pgs}{\PsiT} }
\\
& = \frac{ \int
           \mathcal{D}(\{\vec{x}_m,\vec{y}_n\}) \,
           P(\{\vec{x}_m,\vec{y}_n\}) \,
           \ME{\PsiT}{\prod_m \Bop(\vec{x}_m) \,\Aop\,
                      \prod_n \Bop(\vec{y}_n)}{\PsiT} }
         { \int
           \mathcal{D}(\{\vec{x}_m,\vec{y}_n\}) \,
           P(\{\vec{x}_m,\vec{y}_n\}) \,
           \ME{\PsiT}{\prod_m \Bop(\vec{x}_m) \,
                      \prod_n \Bop(\vec{y}_n)}{\PsiT} }
\\
& = \frac{ \int
           \mathcal{D}(\{\vec{x}_m,\vec{y}_n\}) \,
           P(\{\vec{x}_m,\vec{y}_n\}) \,
           \braket{\eta(\{\vec{x}_m\})}{\phi(\{\vec{y}_n\})} \,
           {\displaystyle
           \frac{\ME{\eta(\{\vec{x}_m\})}{\,\Aop\,}{\phi(\{\vec{y}_n\})}}
                {\braket{\eta(\{\vec{x}_m\})}{\phi(\{\vec{y}_n\})}} }
           }
         { \int
           \mathcal{D}(\{\vec{x}_m,\vec{y}_n\}) \,
           P(\{\vec{x}_m,\vec{y}_n\}) \,
           \braket{\eta(\{\vec{x}_m\})}{\phi(\{\vec{y}_n\})} }
    \,,
}
\end{widetext}
where
\eq{
    \mathcal{D}(\{\vec{x}_m,\vec{y}_n\})
&   \equiv
    {\textstyle \prod_m d\vec{x}_m}
    {\textstyle \prod_n d\vec{y}_n} \,,
\\
    P(\{\vec{x}_m,\vec{y}_n\})
&   \equiv
    {\textstyle \prod_m p(\vec{x}_m)}
    {\textstyle \prod_n p(\vec{y}_n)} \,,
}
and in the last line we have introduced the shorthand
\eq{
    \bra{\eta(\{\vec{x}_n\})}
&   \equiv
    {\textstyle \bra{\PsiT} \prod_n \Bop(\vec{x}_n)} \,;
    \\
    \ket{\phi(\{\vec{y}_m\})}
&   \equiv
    {\textstyle \prod_m \Bop(\vec{y}_m)\ket{\PsiT}}
    \,.
}
The Metropolis simulation is carried out by sampling the probability density
function defined by the integrand in the denominator.
Given the choice of ${\PsiT}$ in the identical-orbital representation,
this readily applies to bosons, which is how the model calculation by
Sugiyama and Koonin\Cite{Koonin1986} was done.
The total length of the imaginary time is predetermined by $\Dt$ and the
number of $\Bop$ operators in the product.

\section{New method for bosons}\label{sec:rnd-wlk} 
\label{ssec:BRW}

In this paper we formulate a new approach for ground-state calculations
of bosons with branching random walks.
There are several advantages in implementing the Monte Carlo sampling as a
random walk process. It is a true ground-state formalisms with open-ended
random walks which allow projection to long enough imaginary-times. The
sampling process can be made much more efficient than in standard AFQMC,
by virtue of importance sampling with ${\PsiT}$ to guide the random walks.
It also leads to a universal approach for bosons and fermions, where it is
necessary to use the random walk formalism in order to implement a
constraint to deal with the sign and complex-phase
problems\Cite{Zhang1997_CPMC,Zhang2003}.

A key observation is that we can choose an IOR single-permanent wave
function as the initial wave function $\ket{\PsiT}$.
At each imaginary timestep $\tau\equiv n\,\Dt$ in the projection in
Eq.~\eqref{eq:gsProjection}, the wave function is stochastically sampled
by a collection of single-permanent wave functions $\{\Wlkr{i}{\tau}\}$,
where the index $\wlkridx{i}$ (in Cursive letter) is different from the
basis index $i$.
From Eqs.~\eqref{eq:Pgs-generic} and \eqref{eq:exp-operator}, we see that,
with each walker $\Wlkr{i}{0}$ initialized to $\ket{\PsiT}$ in IOR, the
resulting projection will lead to a superposition of single-permanent wave
functions, all of which are in IOR.

Each permanent evolves by the stochastic application of $\Pgs$, as
follows: we randomly sample $\vec{x}$ from the probability density
function $p(\vec{x})$, then apply $\Bop(\vec{x})$ on $\Wlkr{i}{\tau}$:
\eql{eq:Pgs-applied}
{
    \Wlkr{i}{\tau+\Dt} \leftarrow
    \Bop(\vec{x}) \Wlkr{i}{\tau} \,,
}
We will call these permanents \emph{random walkers}.
The collection of these random walkers at each imaginary-time step
is also referred to as \emph{population}.

The population must first be equilibrated so that the ground-state
distribution is reached. After equilibrium the ground state is given
stochastically by the collection of single-permanent wave functions
$\{ \Wlkr{i}{} \}$:
\eql{eq:mc-gswf}
{
    \ket{\PsiGS} \doteq \sum_{\wlkridx{i}} \Wlkr{i}{} \,.
}
Measurement of ground-state observables can then be carried out.

The random walk process naturally causes the walker's orbitals to
fluctuate. In order to increase sampling efficiency, we may associate
a \emph{weight} factor $\Ovlp{i}{}$ to each walker $\Wlkr{i}{}$.
For example, we can use the walker's amplitude as the weight factor:
\eq{
    \Ovlp{i}{} \equiv \sqrt{\braket{\wlkr{i}{}}{\wlkr{i}{}}} \,.
}
A better definition of the weight will be introduced later when we discuss
importance sampling.
We duplicate a walker when its weight exceeds a preset
threshold. Conversely, walkers with small weight (lower than a
predetermined limit) should be removed with the corresponding probability.
In this way, the walkers will have roughly the same weight.
This results in a branching random walk.

\subsection{Measurement: ``brute force'' and mixed estimators}

The ground-state value of an observable $\Aop$ is its
expectation value with the ground-state wave function:
\eql{eq:g.s.obs}
{
    \langle \Aop \rangle_{\textrm{g.s.}}
  = \frac{\ME{\PsiGS}{\Aop}{\PsiGS}}{\braket{\PsiGS}{\PsiGS}} \,.
}
In principle, we can use the same Monte Carlo samples as both
$\bra{\PsiGS}$ and $\ket{\PsiGS}$.
A ``brute force'' measurement on population $\{ \Wlkr{i}{\tau} \}$ at
imaginary-time $\tau$ is then given by
\eql{eq:brute-snapshot}
{
    \bruteXV{\Aop}^{(\tau)}
    \equiv
    \frac{\sum_{\wlkridx{i}\wlkridx{j}} 
          \ME{\wlkr{j}{\tau}}{\Aop}{\wlkr{i}{\tau}}}
         {\sum_{\wlkridx{i}\wlkridx{j}} 
          \braket{\wlkr{j}{\tau}}{\wlkr{i}{\tau}}}
}
and the estimator $\bruteXV{\Aop}$ is the average of such measurements.
The ``brute force'' estimator is not useful in real-space based QMC
methods such as diffusion Monte Carlo, because the overlaps between
different walkers would lead to $\delta$-functions.
Here the walkers are non-orthogonal mean-field wave functions, and
Eq.~\eqref{eq:brute-snapshot} is well defined in principle.
The estimator is exact for all observables in the limit of large $\Nwlkr$.
The ground-state energy estimated in this way is variational, namely, the
computed energy lies higher than the exact value (outside of the
statistical errorbar) and converges to the exact value as $\Nwlkr$
is increased.
In practice, however, the usefulness of the `brute force'' estimator is
limited to smaller systems. In general it will have large variances.
Reducing the variance is expensive because $\bruteXV{\Aop}$ scales as
$\Order(\Nwlkr^2)$, where $\Nwlkr$ is the size of the population used to
represent $\ket{\PsiGS}$.


The simplest approach to measuring the observables is the
mixed estimator, i.e.\EXTRA{~by replacing $\bra{\PsiGS}$ with
the best guess of the ground state, $\bra{\psiT}$:}{}
\eql{eq:mixed-est}
{
    \mixedXV{\Aop}
  = \frac{\ME{\psiT}{\Aop}{\PsiGS}}{\braket{\psiT}{\PsiGS}} \,.
}
For example, to compute the ground-state energy, we can introduce the
so-called local energy $\EL[\psiT, \phi]$:
\eql{eq:local-energy}
{
    \EL[\psiT, \phi]
& = \frac{\ME{\psiT}{\Hop}{\phi}}{\braket{\psiT}{\phi}}
}
The ground state energy is obtained from the weighted sum of the local
energies associated with each walker:
\eql{eq:g.s.energy1}{
    \mixedEst{E}
  = \frac{\sum_{\wlkridx{i}} \braket{\psiT}{\wlkr{i}{}}
                             \EL[\psiT, \wlkr{i}{}]}
         {\sum_{\wlkridx{i}} \braket{\psiT}{\wlkr{i}{}}}
}
The local energy for each walker can be computed using the formula
given in Appendix \ref{app:IOR}.

The mixed estimator in \Eq{eq:mixed-est} is exact only if the operator
$\Aop$ commutes with the Hamiltonian. Otherwise, a systematic error arises.
Nonetheless the mixed estimator often gives an improvement over the purely
variational estimator:
\eql{eq:psiT-est}
{
    \trialXV{\Aop}
    \equiv
    \frac{\ME{\PsiT}{\Aop}{\PsiT}}
         {\braket{\PsiT}{\PsiT}}
    \,.
}
Two formulas are often employed to correct for the systematic error:
\eqsl{eq:mixed-extrap1}
{
    \XV{\Aop}_{\textrm{extrap1}}
&   \equiv
    2 \mixedXV{\Aop} - \trialXV{\Aop}
    \,;
\\
    \label{eq:mixed-extrap2}
    \XV{\Aop}_{\textrm{extrap2}}
&   \equiv
    \frac{\mixedXV{\Aop}^2}{\trialXV{\Aop}}
    \,.
}
The second formula is useful for quantities such as density profile, where
it must be nonnegative everywhere. These corrections are good only if
$\ket{\PsiT}$ does not differ significantly from $\ket{\PsiGS}$.
In general, we need the back-propagation scheme to recover the
correct ground-state properties. We will describe this method after
introducing importance sampling.

\subsection{Importance sampling}\label{ssec:ImpSamp}
{

In practice, the efficiency of the bare random walk described earlier is
very low, because the random walks ``randomly'' sample
the Hilbert space, and the weights of the walkers fluctuate greatly.
This results in large statistical noise. We formulate an importance
sampling procedure\Cite{Zhang1997_CPMC,Zhang2003}---using the information
provided by the trial wave function $\ket{\PsiT}$---to guide the random
walk into the region where the expected contribution to the wave function
is large.

\subsubsection{Importance-sampled random walkers}

An importance-sampled walker also consists of a permanent and a weight,
although the weight will be redefined according to the projected overlap of the
permanent with the trial wave function.
The purpose is to define a random walk process which will lead to a
stochastic representation of the
ground-state wave function in the form
\eql{eq:z-gswf-def}
{
    \ket{\PsiGS}
    \doteq
    \sum_{\wlkridx{i}{}}
    \zOvlp{i}{} \frac{\Wlkr{i}{}}{\braket{\PsiT}{\wlkr{i}{}}}
    \,,
}
where $\zOvlp{i}{}$ is the new weight of the walker.
The overlap enters to redefine the weight factor such that walkers which
have large overlap with $\ket{\PsiT}$ will be considered ``important'' and
will tend to be sampled more. Such walkers will also have greater
contributions in the measured observables.
Since the permanent now appears as a ratio
${\Wlkr{i}{}} \, / \, {\braket{\PsiT}{\wlkr{i}{}}}$,
its normalization is no longer relevant and can be discarded,
unlike in the unguided random walk.
The only meaningful information in $\Wlkr{i}{}$ is its position in
the permanent space.

\subsubsection{Modified auxiliary-field transformation}

Now we describe the random walk process for the modified walkers.
The goal is to modify $\Pgs$ in Eq.~\eqref{eq:Pgs-generic} such that the
random walk process leads to random walkers with the characteristics
described above in \Eq{eq:z-gswf-def}.
The basic idea is the same as that in Ref.~\onlinecite{Zhang1997_CPMC}.
The main difference is that here we are dealing with bosons. In addition
the HS fields in Ref.~\onlinecite{Zhang1997_CPMC} are discrete Ising-like,
which allowed simplifications in the importance sampling, while here the
auxiliary fields are continuous and thus a more general formalism will be
developed.
Our mathematical derivation here follows that of
Ref.~\onlinecite{Zhang2003}. Up to now we have assumed that
$\braket{\PsiT}{\wlkr{i}{}}$ is real and positive. There is therefore no
additional subtlety with the meaning of importance sampling and the
correct form of the overlap to use, which Ref.~\onlinecite{Zhang2003}
addressed in the context of fermionic calculations with general
interactions.

To derive the importance-sampled propagator, we plug \Eq{eq:z-gswf-def}
into \Eq{eq:gsProjection-2}. We will focus on the two-body propagator, which is
evaluated stochastically and is therefore affected by importance sampling
in a non-trivial way.

The modified propagator, $\zPgs$, consists of two parts.
The first part is the transformation introduced in
\Eq{eq:HSxformn}, which we now rewrite in the following form:
\eql{eq:HSxformn-alt}
{
    \expP{\half\Dt \hat{v}^2}
& = \frac{1}{\sqrt{2\pi}} \int_{-\infty}^{\infty} \! dx \,
    \expP{-\half x^2}
    \expP{\xbar x - \half\xbar^2}
    \expB{\sqrt{\Dt}\,(x - \xbar)\hat{v}}
    \,,
}
where we have added an arbitrary shift $\xbar$ to the auxiliary
field $x$ in the auxiliary-field operator.
This is a change of variable in the integral on the right-hand side and
does not alter the result of the integral.
\EXTRA{(The reweighting by the extra prefactor
$\expP{\xbar x - \half\xbar^2}$ compensates the bias introduced by
$\Vxbar$, therefore we still effectively sample from the original
distribution.)}{}
The new propagator $\zPgs$
must preserve the representation of $\ket{\PsiGS}$
in the form of \Eq{eq:z-gswf-def}; this dictates that the walkers
propagate in the following manner:
\eql{eq:gsProjection-2-modified}
{
    \zOvlp{i}{\tau+\Dt}
    \frac{\Wlkr{i}{\tau+\Dt}}
         {\braket{\PsiT}{\wlkr{i}{\tau+\Dt}}}
    \; {\longleftarrow} \;
    \zOvlp{i}{\tau}
    \frac{\Wlkr{i}{\tau}}
         {\braket{\PsiT}{\wlkr{i}{\tau}}}
    \,.
}
From this requirement comes the second part of the modified propagator,
which is the overlap ratio
%
%
$
    \braket{\PsiT}{\wlkr{i}{\tau+\Dt}} \, / \,
    \braket{\PsiT}{\wlkr{i}{\tau}}
$.
This factor is obtained by bringing the term
$\braket{\PsiT}{\wlkr{i}{\tau+\Dt}}$ in \Eq{eq:gsProjection-2-modified} to
the right-hand side.
It depends on $\ket{\PsiT}$ and the specific path in
auxiliary-field space, and will ``guide'' the random-walk toward the
region where $\braket{\PsiT}{\wlkr{i}{}}$ is large.

Combining the two parts gives an importance-sampled propagator of the form
\eql{eq:z-Pgs}
{
    \zPgs[\wlkr{}{}]
    \approx
    \int d\vec{x} \, p(\vec{x})
    W(\vec{x},\wlkr{}{}) \Bop(\vec{x} - \Vxbar)
    \,,
}
where
\eql{eq:sz-prop-weight}    
{
    W(\vec{x}, \wlkr{}{})
    \equiv
    \frac{ \ME{\PsiT}{\Bop(\vec{x} - \Vxbar)}{\wlkr{}{}} }
         { \braket{\PsiT}{\wlkr{}{}} }
    \,
    \expP{\Vxbar \cdot \vec{x} - \half \Vxbar\cdot \Vxbar}
}
is the aggregate of all the scalar prefactors in the modified propagator.
This propagator takes 
$
    \{ \zOvlp{i}{\tau}, \Wlkr{i}{\tau} \}
$
and advances the population to
$
    \{ \zOvlp{i}{\tau+\Dt}, \Wlkr{i}{\tau+\Dt} \}
$,
both of which represent $\ket{\PsiGS}$ in the form of
\Eq{eq:z-gswf-def}.

Monte Carlo sampling of the new propagator $\zPgs$ is similar to
the one without importance samping.
We sample $\vec{x}$ from a normal Gaussian distribution,
and apply the operator $\Bop(\vec{x} - \Vxbar)$ to the current
walker $\Wlkr{i}{\tau}$. But now we accumulate an extra multiplicative
weight factor $W(\vec{x},\wlkr{i}{\tau})$ every time we apply
\Eq{eq:z-Pgs}:
\eqssl{eq:sz-prop}
{
    \Wlkr{i}{\tau+\Dt}
&   \leftarrow
    \Bop(\vec{x} - \Vxbar) \Wlkr{i}{\tau}
    \label{eq:sz-prop-permanent}
\\
    \zOvlp{i}{\tau+\Dt}
&   \leftarrow
    W(\vec{x},\wlkr{i}{\tau}) \, \zOvlp{i}{\tau}
    \label{eq:sz-prop-label}
    \,.
}
Here we use the customary notation of vector dot product, e.g.
$\Vxbar \cdot \vec{x} \equiv \sum_i \xbar[i] x_i$.
Note that the weight factor $W(\vec{x},\wlkr{i}{\tau})$ depends on both
the current $\big(\wlkr{i}{\tau}\big)$ and future
$\big(\wlkr{i}{\tau+\Dt}\big)$ walker positions.

\subsubsection{The optimal choice for auxiliary-field shift $\Vxbar$}

The optimal importance sampling is achieved when each random walker
contributes equally to the estimator. We therefore choose $\Vxbar$ to
minimize the fluctuation in the weight factor $\zOvlp{i}{}$.
The fluctuation in $\zOvlp{i}{}$ will be minimized if we minimize the
fluctuation in the prefactor Eq.~\eqref{eq:sz-prop-weight}. We do so by
requiring the partial derivatives of this prefactor to vanish with respect
to $x_i$ at its average ($x_i = 0$):
\eq{
    \left.
    \frac{\partial}{\partial x_i} \left[
    \frac{ \ME{\PsiT}{\Bop(\vec{x} - \Vxbar)}{\wlkr{i}{}} }
         { \braket{\PsiT}{\wlkr{i}{}} }
    \times
    \expP{\Vxbar \cdot \vec{x} - \half \Vxbar\cdot \Vxbar}
    \right]
    \right|_{x_i = 0} = 0 \,.
}
It is sufficient to expand the exponentials in terms of $\Dt$ and require
the term linear in $x_i$ to vanish, since this is the leading term,
containing $\sqrt{\Dt}$. The others contain higher-order terms and are
vanishingly small as $\Dt \rightarrow 0$. The best choice for $\xbar[i]$
that satisfies this requirement is
\eql{eq:best-xbar}
{
    \xbar[i] = -\sqrt{\Dt} \,
               \frac{ \ME{\PsiT}{\hat{v}_i}{\wlkr{i}{}} }
                    { \braket{\PsiT}{\wlkr{i}{}} }
        \equiv -\sqrt{\Dt} \, \vbar[i] \,.
}
This choice depends on the current walker position as well as
$\ket{\PsiT}$, which is to be expected, since the objective for the shift
is to guide the random walk toward the region where
$\braket{\PsiT}{\wlkr{i}{\tau}}$ is large.
With $\Vxbar$ determined, the algorithm for the random walk, as given in
Eq.~\eqref{eq:sz-prop}, is now completely specified.

\subsubsection{Local energy approximation}

We can furthermore approximate the prefactor $W(\vec{x},\wlkr{}{})$ in
Eq.~\eqref{eq:sz-prop-weight} to obtain a more elegant and compact
expression.
After rewriting the prefactor in the form of an exponential, expanding
$\Bop(\vec{x} - \Vxbar)$ in terms of $\Dt$, and ignoring terms higher than
$\Order(\Dt)$ in the exponent, we obtain
\EXTRA{~(These are terms of order $\Dt^{3/2}$ and higher in the exponent,
which should contribute $\Order(\Dt^3)$ and higher to the Trotter error
when the integration over the auxiliary fields is performed.)}{}
\eql{eq:sz-prop-weight-toward-EL}
{
    \prod_i
    \expB{\half \Dt (1 - x_i^2) (\vbar[i]^2 - \vsqrbar[i])}
    \expB{\half \Dt \vsqrbar[i]}
\,,
}
where
\eql{eq:vbar2-def}
{
    \vsqrbar[i] \equiv \,
    \frac{ \ME{\PsiT}{\hat{v}_i^2}{\wlkr{i}{}} }
         { \braket{\PsiT}{\wlkr{i}{}} } \,.
}
The product is over the basis index $i$, which should be distinguished
from the walker index $\wlkridx{i}$. The latter is held fixed here.
The first exponential in Eq.~\eqref{eq:sz-prop-weight-toward-EL} can be
ignored by noting that the average value of $x_i^2$ with respect to the
Gaussian probability density function is unity.
Setting $x_i^2 \rightarrow 1$, i.e., evaluating the exponential at the
mean value $\XV{x_i^2}$, is justified because
$\vbar[i]^2$ and $\vsqrbar[i]$ do not change drastically
within one timestep.
We also note that $\sum_i \vsqrbar[i] =
-\ME{\PsiT}{\Vop}{\wlkr{i}{}} / \braket{\PsiT}{\wlkr{i}{}}$, which is
the mixed-estimator of the potential energy with respect to the walker
$\Wlkr{i}{}$.
Combining this term with the similar contribution from the kinetic
propagator, we obtain a simple, approximate expression for
Eq.~\eqref{eq:sz-prop-weight}:
\eql{eq:sz-prop-weight2}
{
    W(\vec{x},\wlkr{i}{\tau})
&   \approx
    e^{\Dt (\ET - \EL[\PsiT, \wlkr{i}{}])}
    \,,
}
where $\EL[\PsiT, \wlkr{i}{}]$ is the local energy of $\wlkr{i}{}$ as
defined in Eq.~\eqref{eq:local-energy}.
Note that, contrary to Eq.~\eqref{eq:sz-prop-weight}, this form depends
only on the current walker position and not the future, although in
practice a symmetrized version can be used which replaces the local energy
by the average of the two. For a good trial wave function, the local
energy fluctuates less in the random walk. If the trial wave function
is the exact ground-state wave function, the local energy becomes a
constant and the weight fluctuation is altogether eliminated.
This bears a close formal resemblance to the importance-sampled difussion
Monte Carlo method.

The algorithm resulting from Eq.~\eqref{eq:sz-prop-weight2} is an
\emph{alternative} to Eq.~\eqref{eq:sz-prop-weight}. The two are identical
and exact in the limit $\Dt \rightarrow 0$, but can have
different Trotter errors.

}

\subsection{Measurement: back propagation}\label{ssec:backprop}

With importance sampling, the mixed estimator in \Eq{eq:mixed-est}
is given by:
\eql{eq:mixed-est-imp}
{
    \mixedXV{\Aop}
  = \frac{ \displaystyle
           \sum_{\wlkridx{i}}
           \zOvlp{i}{}
           \frac{\ME{\PsiT}{\Aop}{\wlkr{i}{}}}
                {\braket{\PsiT}{\wlkr{i}{}}} }
         { \displaystyle
           \sum_{\wlkridx{i}} \zOvlp{i}{} }
    \,.
}
For example, the ground-state energy is
\eq{
    \mixedEst{E}
  = \frac{ \sum_{\wlkridx{i}} \zOvlp{i}{} \EL[\psiT, \wlkr{i}{}] }
         { \sum_{\wlkridx{i}} \zOvlp{i}{} }
    \,.
}
As mentioned earlier, the normalization of $\wlkr{i}{}$ is irrelevant
because $\wlkr{i}{}$ only appears in ratios in any formula that defines
the algorithm: \Eqs{eq:z-gswf-def}, \eqref{eq:sz-prop-weight},
\eqref{eq:best-xbar}, \eqref{eq:sz-prop-weight2}, and
\Eq{eq:mixed-est-imp}. We can (and should) normalize the permanent as
needed, and discard the resulting normalization factor.

The mixed estimator is often inadequte
for computing observables whose operators do
not commute with the Hamiltonian. In some cases the error due to this
noncommutation is unacceptable. For example, the condensate fraction in
the attractive trapped Bose-Hubbard model is greater than 100\% if
the Green's function $\langle \Cc{i}\Dc{j} \rangle$ is estimated using the
mixed estimator. Therefore we have to propagate the wave functions on both
the right- and the left-hand side of the operator:
\eql{eq:bp-obs-def}
{
    \BPEst{\XV{\Aop}}
  = \frac{\ME{\PsiT}{\exptBPH\Aop}{\PsiGS}}
         {\ME{\PsiT}{\exptBPH}{\PsiGS}} \,.
}
This estimator approaches the exact expectation value in
Eq.~\eqref{eq:g.s.obs} as $\tauBP$ is increased.
Zhang and co-workers proposed a back-propagation
technique\Cite{Zhang1997_CPMC} that reuses the auxiliary-field ``paths''
from different segments of the simulation to obtain
$\bra{\PsiGSBP} \equiv \bra{\PsiT}\exptBPH$,
while avoiding the $\Nwlkr^2$ scaling of a brute-force evaluation with
two separate populations for $\bra{\PsiGS}$ and $\ket{\PsiGS}$.
Here we give a more formal derivation and description of the technique,
and implement it to bosons.

At imaginary-time $\tau$, the population is $\{\Wlkr{i}{\tau}\}$,
which represents $\ket{\PsiGS}$ in the form of \Eq{eq:z-gswf-def}.
The propagator in the denominator can be viewed equivalently as operating
on the left or the right.
The latter view is precisely the ``normal'' importance-sampled random walk
from $\tau$ to the future time $\tau' \equiv \tau + \tauBP$, which
consists of $\nBP \equiv \tauBP / \Dt$ steps.
We first assume that there is no branching (birth/death of
walkers), i.e., the weights are fully multiplied according to
\Eq{eq:sz-prop-weight}.
\EXTRA{%
(It is straightforward to generalize to the case with branching,
which we will do shortly.)
}{}%
The random walk of each walker will generate a path in auxiliary-field
space.
For convenience we will denote the path-dependent operator
$\Bop[\vec{x}_{\wlkridx{i}}^{(\tau)} - \Vxbar(\wlkr{i}{\tau})]$ by
$\bpath{i}{\tau}$, and weight factor
$W(\vec{x}_i^{(\tau)},\wlkr{i}{\tau})$ by
$\wpath{i}{\tau}$.
Further we will denote the time-ordered product of $\bpath{i}{\tau}$ from
imaginary-time $\tau$ to $\tau'$ by $\bpath{i}{\tau':\tau}$, and
correspondingly the product of $\wpath{i}{\tau}$ by
$\wpath{i}{\tau':\tau}$.
Each path defines a product
\eql{eq:z-path-def}
{
    \frac{ 1 }{ \braket{\PsiT}{\wlkr{i}{\tau'}} }
    \wpath{i}{\tau':\tau} \bpath{i}{\tau':\tau}
    { \braket{\PsiT}{\wlkr{i}{\tau}} }
    \,.
}
Collectively these products give a stochastic representation of
$\exptBPH$.

Replacing the operator $\exptBPH$ in the numerator and denominator of
\Eq{eq:bp-obs-def} with \Eq{eq:z-path-def}, and using the
expression for $\ket{\PsiGS}$ given by \Eq{eq:z-gswf-def}, we obtain
\eql{eq:bp-est-a}
{
    \BPXV{\Aop}
& = \frac{ 
           \sum_{\wlkridx{i}}
           {\bra{\PsiT} \,
           \frac{1}{\braket{\PsiT}{\wlkr{i}{\tau'}}}
           \wpath{i}{\tau':\tau} \bpath{i}{\tau':\tau} \Aop \,
           \zOvlp{i}{\tau} \Wlkr{i}{\tau}}
         }
         { 
           \sum_{\wlkridx{i}}
           {\bra{\PsiT} \,
           \frac{1}{\braket{\PsiT}{\wlkr{i}{\tau'}}}
           \wpath{i}{\tau':\tau} \bpath{i}{\tau':\tau}
           \zOvlp{i}{\tau} \Wlkr{i}{\tau}}
         }\,.
}
Using the propagation relation in Eq.~\eqref{eq:sz-prop},
we can show that
\eql{eq:bp-denom}
{
    \bpath{i}{\tau':\tau} W_\wlkridx{i}^{({\tau':\tau})}\;
    \zOvlp{i}{\tau}
    {\Wlkr{i}{\tau}}
  = \zOvlp{i}{\tau'}
    {\Wlkr{i}{\tau'}}
    \,,
}
i.e., the denominator in Eq.~\eqref{eq:bp-est-a} reduces to
$\sum_{\wlkridx{i}} \zOvlp{i}{\tau'}$. This result is to be expected, and
can also be seen by completing the $\nBP$ steps of the ``normal'' random
walk we discussed above. With importance sampling, the Monte Carlo
estimate of the denominator
is simply given by the weights at time $\tau'$.

To simplify the numerator we associate a \emph{back-propagated} wave
function with each walker $\Wlkr{i}{\tau}$
\eql{eq:backprop-wf-def}
{
    \bWlkr{i}{\tauBP}
    \equiv
    \left[\bpath{i}{\tau+\tauBP~:~\tau}\right]^\dagger \ket{\PsiT} \,.
}
Note that each of these $\eta$'s originates from the trial wave function
$\ket{\PsiT}$, and is propagated by applying the $\Bop$'s in
\emph{reverse} order, as implied by the Hermitian conjugation.
We may then write Eq.~\eqref{eq:bp-est-a} in the following form:
\eql{eq:bp-est}
{
    \BPXV{\Aop}
& = \frac{ \displaystyle
           \sum_{\wlkridx{i}} \zOvlp{i}{\tau'}
           \frac{ \ME{\bwlkr{i}{\tauBP}}{\Aop}{\wlkr{i}{\tau}} }
                { \braket{\bwlkr{i}{\tauBP}}{\wlkr{i}{\tau}} } }
         { \displaystyle
           \sum_{\wlkridx{i}} \zOvlp{i}{\tau'} }
    \,.
}

The estimators in \Eqs{eq:bp-obs-def} and \eqref{eq:bp-est}
parallel that of the standard AFQMC estimator in Eq.~\eqref{eq:MMC-AFQMC}.
The $\ket{\phi}$'s and $\bra{\eta}$'s have similar meanings.
The only difference lies in how the paths are generated.
Here an open-ended random walk is used to advance an ensemble of paths
from $\tau$ to $\tau'$, which result in fluctuating weights that represent
the path distribution. In standard AFQMC a fixed length path
(corresponding to $\tauBP + \tau_{\rm eq}$, with $\tau_{\rm eq}$ being the
minimum time for equilibriation or, failing that, the maximum time that
can be managed by the calculation) is moved about by the Metropolis
algorithm, which eliminates branching by the acceptance/rejection step. In
other words, the estimators in \Eq{eq:MMC-AFQMC} and \Eq{eq:bp-est} are
the same except for the weights.

Eq.~\eqref{eq:bp-est} defines an algorithm for obtaining the estimate of
$\BPXV{\Aop}$ via the following steps:
\begin{enumerate}
\item A population is recorded as $\{ \Wlkr{i}{\tau} \}$;
\item as the random walk continues, the path history is kept
for a time interval $\tauBP$;
\item the population $\{ \bWlkr{i}{\tauBP} \}$ is then generated
by back-propagation using Eq.~\eqref{eq:backprop-wf-def};
\item this population is matched in a one-to-one manner to
$\{ \Wlkr{i}{\tau} \}$, weighted by the weight \emph{at the later
time}, $\zOvlp{i}{\tau'}$, and the estimator is formed.
\end{enumerate}
In the back-propagation the propagators are, as shown in
\Eq{eq:backprop-wf-def}, idential to those in the forward direction, but
in reverse order in imaginary-time. As in the normal walk, the normalization
of $\bWlkr{i}{\tauBP}$ does not enter in the estimator.
Similar to the mixed estimator, this procedure can be repeated
periodically to improve statistics.
Evidently this estimator is exact in the limit of large $\tauBP$.

We have assumed that there is no branching within the interval $\tauBP$.
In practice, a population control scheme is often used which causes
birth/death of walkers. This does not affect the derivation above or the
basic algorithm. The effect on the implementation is that a list of
ancestry links must be kept for the forward steps, which indicates the
parent of each walker at each step in the imaginary-time duration
$\tauBP$.
As a result of branching, two or more $\bra{\eta}$'s may share the same
segment of the paths in their ``past'' and the same parent
$\Wlkr{i}{\tau}$. The estimator remains exact for large $\tauBP$.
Branching or weight fluctuation does have a more serious
practical implication, however. As $\tauBP$ is increased, more and more
$\bra{\eta}$'s will be traced back to the same parent $\Wlkr{i}{\tau}$. Or
equivalently, fewer and fewer permanents in the set $\{\Wlkr{i}{\tau}\}$
will contribute to the estimator. This results in a loss of efficiency or
an increase in variance. Better importance sampling will help improve the
situation, often greatly, by reducing fluctuations in weights, although the
problem will always occur at large enough $\tauBP$. In our applications to
date we have
rarely encountered the problem and find that the computed observables
converge quite rapidly
(see section \ref{sec:DemoResults} for illustrative results).


\section{Trapped boson gas: model and implementations of
QMC and GP methods} 
\label{sec:TBH}

In this section we discuss the model we use to describe a single-species,
Bose atomic gas with pair-wise contact interaction, confined in a harmonic
trap in one- or three-dimensions. We then describe the implementations of
both our QMC method and the standard mean-field GP approach to study this
model.
Numerical results will be presented in the following section,
\Sec{sec:DemoResults}.

\subsection{Model}

We use an effective potential characterized by the low energy atom-atom
scattering length, $a_s$. The two-body interaction takes a simple form
\eql{eq:2B-potential}
{
    U(\mathbf{r}_1 - \mathbf{r}_2)
  = \frac{4\pi a_s \hbar^2}{m} \delta(\mathbf{r}_1 - \mathbf{r}_2) \,.
}
For this effective potential to be valid, several assumptions are made;
for example, the dominant effect is from $s$-wave scattering, and
$|a_s|$ is much smaller than the average inter-particle spacing.
For more details we refer the reader to Ref.~\onlinecite{Leggett01}.
In the alkali gases these conditions are in general well met,
and the model potential can be expected to give quatitative information,
although care must be taken to validate the conditions.

We now derive the Bose-Hubbard model from the standard many-body
Hamiltonian of the trapped boson problem in $d$-dimension. In the
continuous, real space, the Hamiltonian is given by:
\begin{widetext}
\eql{eq:Hamiltonian_real}
{
    \Hop
  = \Kop + \Vop
& = \spinsum{\sigma}
    \intdr \Cpsi{\spin{\sigma}}(\rvec) \,
    \left(-\frac{\hbar^2}{2m}\nabla_\rvec^2 + \half m\omega_0^2 r^2\right)
    \Dpsi{\spin{\sigma}}(\rvec)
    \\
& +~\Half \cdot \frac{4\pi a_s \hbar^2}{m}
    \spinsum{\sigma}
    \intdr[\_1] \intdr[\_2]
    \Cpsi{\spin{\sigma}}(\rvec\_1) \Cpsi{\spin{\sigma}}(\rvec\_2)
    \delta(\rvec\_1 - \rvec\_2)
    \Dpsi{\spin{\sigma}}(\rvec\_2) \Dpsi{\spin{\sigma}}(\rvec\_1) \,.
}
\end{widetext}
The first term is the one-body Hamiltonian $\Kop$, which consists of the
kinetic energy and the (external) confinement potential. $\Vop$ is the
interaction Hamiltonian, which is the sum of all the two-body potentials.
The characteristic trap frequency is $\omega_0$, which is related to the
so-called {oscillator length scale} by
$\aho = \sqrt{\hbar/m\omega\_0}$.

We introduce a real-space lattice, with a linear dimension of $L$,
in a simulation cell of volume $(2r_b)^d$.
The lattice spacing is therefore $\varsigma = 2r_b / L$. Further we will
consider only a spherically symmetric trap here for simplicity.
We truncate the simulation cell accordingly and assume that the wave
function is negligible outside the maximum sphere enclosed by the cell.
(Generalization to inhomogeneous traps is straightforward.)

The discretized Hamiltonian corresponding to \Eq{eq:Hamiltonian_real} is
\begin{widetext}
\eql{eq:Hubbard_H_final}
{
    \Hop
& = \spinsum{\sigma} \sum_i \left\{
  - t \Big[ \sum_{j\in\mathrm{nn}(i)}\!\!
            \Cc{i\spin{\sigma}}\Dc{j\spin{\sigma}}
          - 2d \Cc{i\spin{\sigma}}\Dc{i\spin{\sigma}} \Big]
  + \half \kappa |\tilde\rvec\_i - \tilde\rvec\_0|^2
    \Cc{i\spin{\sigma}}\Dc{i\spin{\sigma}}
    \right\}
  +~\half U \spinsum{\sigma} \sum_i
    \left(\Cc{i\spin{\sigma}} \Dc{i\spin{\sigma}}
          \Cc{i\spin{\sigma}} \Dc{i\spin{\sigma}}
          - \Cc{i\spin{\sigma}} \Dc{i\spin{\sigma}} \right) \,,
}
\end{widetext}
where $\Cc{i}$ and $\Dc{i}$ are the usual creation and annihilation operators
at site $i$.
The Hubbard parameters $t$, $U$, and $\kappa$ are related to the real,
physical parameters as follows:
\eqssl{eq:Hub2real}
{
    t      &= \frac{1}{2\varsigma^2}       \label{eq:Hub2real-t} \\
    U      &= \frac{4\pi a_s}{\varsigma^d} \label{eq:Hub2real-U} \\
    \kappa &= \frac{\varsigma^2}{\aho[4]}  \label{eq:Hub2real-k} \,,
}
where for simplicity we have set $\hbar = m = 1$.
The lattice coordinate $\tilde\rvec\_i$ is related to the real coordinate
by $\tilde\rvec\_i = (L / 2 r_b)\rvec\_i$, and
$\tilde\rvec\_0$ is the lattice coordinate of the trap's center.
Note that $a_s$ is the true scattering length only in three-dimensional
systems. Nonetheless we will retain the symbol $a_s$ in
\Eq{eq:Hub2real-U} as a convenient measure of the interaction strength in
any dimension.

In the discretized model our resolution is limited by the lattice spacing.
This is consistent with the conditions of validity of the model
interaction in \Eq{eq:2B-potential}, as it in a sense ``integrates
out'' the short-range dynamics.
In this model our lattice constant $\zeta$ must be much smaller compared to
the average interparticle spacing, but larger than the scattering
length:
\eql{eq:Veff-reqs-lattice}
{
    |a_s| \ll \zeta \ll \rho^{-1/d} \,.
}
With negative $a_s$, the particles tend to ``lump'' together due to the
gain in the interaction energy. This is a situation where we especially
have to be aware of the validity of the effective potential. As mentioned
we will do a consistency check at the end of the calculation to ensure
that the occupancy of the lattice points are less than unity.

\subsection{Implementation of QMC}
\label{ssec:QMC-implement}

Implementation of our QMC method for this model is straightforward.
The number of basis $M$ is equal to the number of lattice sites inside the
truncated sphere of radius $r_b$.
The two-body term in Eq.~\eqref{eq:Hubbard_H_final} is in the desired form
of Eq.~\eqref{eq:VopSumSqr}. With a negative $U$, the HS transformation in
\Eq{eq:HSxformn-general} leads to $M$ auxiliary fields, with one-body
propagators in the form of $\exp({\sqrt{\Dt |U|} x_i \nop{i}})$, where
$\nop{i} \equiv \Cc{i\spin{\sigma}} \Dc{i\spin{\sigma}}$
is the density operator.
Our trial wave function $\ket{\PsiT}$ is the Gross-Pitaevskii (GP) wave
function $\PsiGP$, which we describe in the next subsection.


We mention here a technical point in the implementation.
The ground-state projection in our method involves the application of
one-body propagator in the form of $e^{\Aop}$ on
a single-permanent wave function $\ket{\phi}$.
This usually translates into a matrix-vector multiplication in the
computer program, which generally costs $\Order(M^2)$%
\EXTRA{---which is expensive for large systems.
Rapid evaluation of $\e^{\Aop}\ket{\phi}$ is therefore desirable}{}.
Often there are special properties of $\Aop$ that can be exploited to
evaluate the one-body propagator more efficiently.
In the Bose-Hubbard Hamiltonian, the only non-diagonal part of the
Hamiltonian in real space is the kinetic operator in $\Kop$.
We can separate it from the other one-body operators and apply
the kinetic propagator in momentum space.
Wave functions are quickly translated between these two representations
using the Fast Fourier transform (FFT).
In this way, the actual application of $\expdtK$ involves only diagonal
matrices; thus the overall cost for each $\expdtK$ operation is
reduced to $\Order(M\log M)$.
We observe in our calculations that the additional Trotter error is much
smaller than the error already introduced in the original breakup,
Eq.~\eqref{eq:Trotter-Suzuki}.

\subsection{Implementation of Gross-Pitaevskii self-consistent equation}
\label{ssec:GP-implement}

The Gross-Pitaevskii (GP) wave function $\PsiGP$ is the single-permanent
wave function
\eql{eq:GP-wf-assumption}
{
    \PsiGP(\rvec_1, \rvec_2, \ldots \rvec_N) =
    \varphi(\rvec_1) \varphi(\rvec_2) \cdots \varphi(\rvec_N) \,,
}
which minimizes the expectation value of the ground-state energy. Such a
wave function satisfies the self-consistent Gross-Pitaevskii
equation\Cite{Gross1961,Gross1963,Pitaevskii1961}
\eql{eq:GP_Hamiltonian_real}
{
& - \frac{\hbar^2}{2m} \nabla^2 \varphi(\rvec)
  + \half m\omega_0^2 |\rvec - \rvec\_0|^2 \varphi(\rvec)
\\
&   \qquad\qquad
  + \frac{N-1}{N}
    \frac{4\pi a_s \hbar^2}{m} |\varphi(\rvec)|^2 \varphi(\rvec)
  = \mu \varphi(\rvec) \,.
}
[We keep the prefactor $(N-1) / N$, since we will study both large and
small values of $N$.]

To compare our QMC results to those of mean-field, we carry out GP
calculations on the same lattice systems.
The discretized GP Hamiltonian in the second-quantized form is:
\eql{eq:meanfield_Hubbard_H}
{
    \Hop\_{\mathrm{GP}}
& = -t \spinsum{\sigma} \sum_i
    \Big( \sum_{j\in\mathrm{nn}(i)}\!\!
          \Cc{i\spin{\sigma}}\Dc{j\spin{\sigma}}
        - 2d \Cc{i\spin{\sigma}}\Dc{i\spin{\sigma}} \Big)
\\
& +~\half\kappa \spinsum{\sigma} \sum_i
    |\tilde\rvec\_i - \tilde\rvec\_0|^2
    \Cc{i\spin{\sigma}}\Dc{i\spin{\sigma}}
\\
& +~\frac{N-1}{N} \, U \spinsum{\sigma} \sum_i
    \left(\nbar\_{i\spin{\sigma}} \Cc{i\spin{\sigma}} \Dc{i\spin{\sigma}}
         - \half\nbar_{i\spin{\sigma}}^2\right) \,.
}
Here $\nbar_{i\spin{\sigma}}$ is the expectation value of the density
operator:
\eql{eq:nbar-def}
{
    \nbar_{i\spin{\sigma}}
    \equiv
    \frac{\ME{\PsiGP}{\Cc{i\spin{\sigma}} \Dc{i\spin{\sigma}}}{\PsiGP}}
         {\braket{\PsiGP}{\PsiGP}} \,.
}
\EXTRA{The complexity of the many-body Hamiltonian is thus alleviated by
introducing an effectively one-body Hamiltonian, which can be diagonalized
at a much lower cost than the full Hamiltonian,
Eq.~\eqref{eq:Hubbard_H_final}.}{}

We have implemented two methods for solving the GP equation.
The \emph{first} is the usual self-consistent iterative approach.
We generate an initial density profile, $\nbar_{i\spin{\sigma}}^{(0)}$,
by solving the non-interacting Hamiltonian (with $U = 0$).
The density is fed back to construct the initial Hamiltonian
$\Hop_{\mathrm{GP}}^{(0)}$ in \eqref{eq:meanfield_Hubbard_H}.
Direct diagonalization of this one-body Hamiltonian yields
its ground state $\ket{\PsiGP[(1)]}$.
We thus obtain an updated density $\nbar_{i\spin{\sigma}}^{(1)}$ and a
better Hamiltonian $\Hop_{\mathrm{GP}}^{(1)}$. This procedure is iterated
until the desired convergence criterion is satisfied.
We choose our convergence condition to be:
\eql{eq:GP-conv-cond}
{
    \frac{\int d\rvec |\varphi^{(t+1)}(\rvec) - \varphi^{(t)}(\rvec)|}
         {\half
          \int d\rvec |\varphi^{(t+1)}(\rvec) + \varphi^{(t)}(\rvec)|}
  < \epsilon \,,
}
where $\epsilon$ is a small number (usually on the order of $10^{-13}$ for
double precision numbers).

The \emph{second} method we use to solve \Eq{eq:meanfield_Hubbard_H}
avoids the diagonalization procedure. It is closely related to the QMC
method, both computationally and formally (see \Sec{sec:Discussions}).
We use the ground-state projector $\expP{-\Dt\Hop\_{\mathrm{GP}}}$:
\eql{eq:GP.proj}
{
    (\expP{-\Dt\Hop\_{\mathrm{GP}}})^n \ket{\Psi^{(0)}}
    \overset{n\rightarrow\infty}\longrightarrow \ket{\PsiGP} \,.
}
The initial wave function is arbitrary and can be, for example, chosen
again as the solution with $U = 0$. The feedback mechanism through the
density profile $\nbar\_{i\spin{\sigma}}$ remains the same.
By using the same Fast Fourier transform for the kinetic propagator as
described in subsection \ref{ssec:QMC-implement}, a speed gain is
obtained, especially for large systems.
In practice we have often found this method to be a simpler and faster
alternative to the first method of diagonalization and iteration.
Note that the scalar term
$
    -\half \frac{N-1}{N} U \sum_i \nbar_{i\spin{\sigma}}^2
$
does not affect the projection process, but with it $\Hop\_{\mathrm{GP}}$
corresponds to the original many-body Hamiltonian in that
$
    \ME{\PsiGP}{\Hop\_{\mathrm{GP}}}{\PsiGP} =
    \ME{\PsiGP}{\Hop}{\PsiGP}
$.

\section{Results} 
\label{sec:DemoResults}

In this section we present results from our QMC and GP calculations in
one-, two-, and three-dimensions.
To validate our new QMC method and illustrate its behavior, the
majority of the calculations will be on systems where exact results are
available for benchmark. These include small lattices, which can be
diagonalized exactly, and the case of attractive $\delta$-function
interactions in one dimension, where analytic solutions exist.
For the purpose of presenting the method to facilitate implementation,
some numerical results and comparisons are shown in detail
to illustrate the behavior and characteristics of the method.

Most of the results we present here will be for attractive interactions,
where the method is exact and is free of any phase problem\Cite{Zhang2003}
from complex propagators (see subsection \ref{ssec:Initial-repulsive}).
Such systems therefore provide a clean testground for our new method.
In addition, with attractive interactions the condensate in 3-D is believed
to collapse beyond a critical interaction strength or number of particles.
Mean-field calculations\Cite{Ruprecht1995} estimate the collapse critical
point to be about $N \as / \aho = -0.575$.
The exact behavior of the condensate near the critical point is, however,
not completely clear, as many-body effects are expected to have an impact.
At the end of this section we will also show some preliminary results for
larger systems with both attractive and repulsive interactions in 3-D.

We measure the ground-state expectation values of the following
quantities: the ground-state energy, kinetic energy $\kinXV$, external
confining potential $\trapXV$, interaction energy $\potXV$, density
profile $\XV{\nop{i\spin{\sigma}}}$, and the condensate fraction (often
abbreviated ``cond.frac.'' in the tables and figures). The {condensate
fraction} is defined as the largest eigenvalue of the diagonalized density
matrix\Cite{Leggett01}.
If we write the one-body Green's function matrix $\XV{\Cc{i}\Dc{j}}$
in terms of its eigenvalues $\{n_\alpha\}$ and
eigenvectors $\{\chi\_\alpha(i)\}$:
\eq{
    \langle{\Cc{i}\Dc{j}}\rangle =
    \sum_{\alpha} n_\alpha \chi_\alpha^\dagger(i) \chi_\alpha(j) \,,
}
then the largest eigenvalue divided by the total number of particles
gives the condensate fraction.

\subsection{Comparison with exact diagonalization: $a_s < 0$} 

The many-body Hamiltonian \eqref{eq:Hubbard_H_final} can be diagonalized
exactly for small systems to benchmark our QMC calculation. We compare our
QMC results with exact diagonalization for a one-dimensional lattice of 13
sites, and study its behavior for different values of the interaction
strength $a_s$ and number of particles $N$.


The first system we study has 5 bosons, with
$t = 2.676$, $U = -1.538$, $\kappa = 0.3503$.
These values correspond to the physical parameters
$\aho = 8546$~\AA{} and $\as = -5.292 \times 10^{-6}$~\AA$^{-1}$.
(Recall that, by our definition, $\as$ in 1-D does not have the dimension
of length, and is not the scattering length itself.)
Table \ref{tbl:T1D-13s5p.Ux1} shows the comparison of the quantities
computed using three methods: QMC, GP, and exact diagonalization (ED).
The statistical uncertainty of QMC results are presented in parantheses.
We see that the agreement between QMC and ED is excellent.
GP makes significant errors here because of the sizable interaction
strength as well as the small number of particles.

\begin{table}[htbp!]
\begin{center}

  \caption{\label{tbl:T1D-13s5p.Ux1}
  Comparison of QMC calculation against exact diagonalization (ED)
  and Gross-Pitaveskii (GP). The system has 13 sites, 5 particles,
  $t = 2.676$, $U = -1.538$, $\kappa = 0.3503$.
  In the QMC calculation we use $\Dt = 0.01$, $\tauBP = 4.0$, and
  the GP solution as the trial wave function.
  }
  \begin{tabular}{llllll}
  \hline
   Type     &  g.s.energy   &  $\kinXV$    &  $\trapXV$   &  $\potXV$     & cond.frac. \\
  \hline
  \hline
   ED       &  $-1.009$     &  $4.278$     & $0.8427$     &  $-6.129$     & $95.59\%$  \\
   QMC      &  $-1.008(2)$  &  $4.279(3)$  & $0.8423(5)$  &  $-6.129(2)$  & $95.59\%$  \\
   GP       &  $-0.493 $    &  $3.919$     & $0.7504$     &  $-5.162$     & $100\%$    \\
  \hline
  \end{tabular}

\end{center}
\end{table}

\begin{figure}[htbp!]

  \includegraphics[scale=0.75]{\PLOTFILE{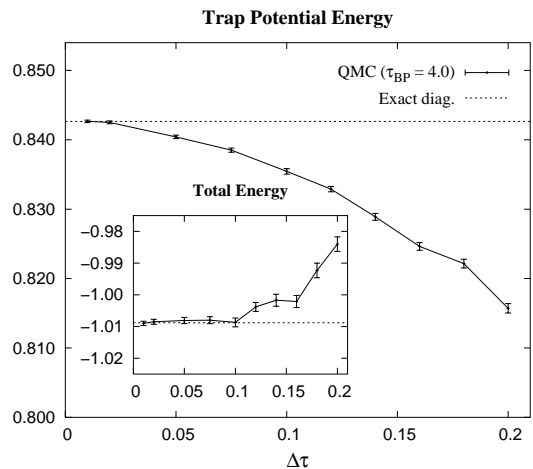}}
  \caption{\label{fig:dt-convergence1}
  Convergence of QMC observables with $\Dt$.
  The system has the same parameters as in Table \ref{tbl:T1D-13s5p.Ux1}.
  Exact results are shown as \PLOTCOLOR{blue}{dotted} lines.
  Lines connecting QMC data are to aid the eye.
  }

\end{figure}

To illustrate the convergence in imaginary-timestep $\Dt$, we show in
\Fig{fig:dt-convergence1} the total energy and the average trap energy
$\trapXV$. The former can be obtained exactly from the mixed estimator
while the latter requires back propagation.
To show the Trotter error, we have deliberately done the calculations up
to rather large $\Dt$ values. We see that both quantities converge to the
exact results as $\Dt \rightarrow 0$.

\begin{figure}[htbp!]

  \includegraphics[scale=0.7]{\PLOTFILE{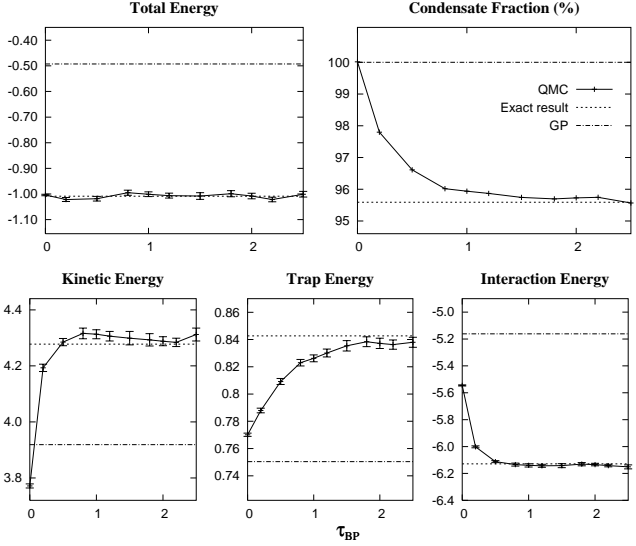}}
  \caption{\label{fig:dtBP-convergence1}
  Convergence of the computed observables versus $\tauBP$.
  The system is the same as in Table \ref{tbl:T1D-13s5p.Ux1}.
  The different panels show five different observables.
  The horizontal axes are the back-propagation length.
  Exact results are shown as \PLOTCOLOR{blue}{dotted} lines, while
  GP results as \PLOTCOLOR{green}{dash-dotted} lines.
  Solid lines are present only to aid the eye.
  }

\end{figure}

To illustrate the convergence of observables in back-propagation length,
we show in \Fig{fig:dtBP-convergence1} the various observables computed by
QMC as a function of $\tauBP$.
Separate calculations were done for different values of $\tauBP$. For all
calculations, a small $\Dt$ value of $0.01$ was used.
We see that all quantities converge to the exact results rather quickly,
by $\tauBP \sim 2$. (The total energy $\langle H\rangle$ is of course
exact for any $\tauBP$, including $\tauBP=0$.)
As we see from the energy expectations, this is in fact a system with
significant interaction effects. Alkali systems at the experimental
parameters often have significantly weaker interaction strengths and the
convergence rate is expected to be even faster.

\begin{figure}[htbp!]

  \includegraphics[scale=0.66]{\PLOTFILE{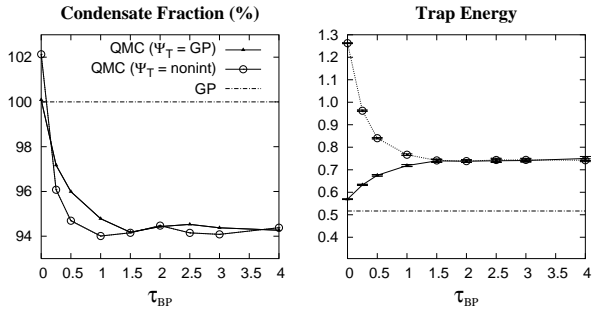}}
  \caption{\label{fig:dtBP-convergence2}
  Independence of QMC results on trial wave functions (``{GP}'' for
  Gross-Pitaevskii, ``{nonint}'' for noninteracting solution).
  The system is the same as in Table \ref{tbl:T1D-13s5p.Ux1},
  except that here we use 6 particles.
  The horizontal axes are the back-propagation length.
  Lines connecting QMC data points are present only to aid the eye.
  }

\end{figure}

Our QMC method is exact and therefore independent of the trial wave
function $\PsiT$, except for convergence rate and statistical errors.
In \Fig{fig:dtBP-convergence2} we show QMC results obtained using two
different $\PsiT$'s, the noninteracting solution and the GP wave function.
The convergence of condensate fraction and trap energy are shown versus
back-propagation time $\tauBP$ for a system of 6 particles on 13 sites.
The calculations lead to the same results.
The quality of $\PsiT$, however, does affect the variances of the
observables and their convergence rates with $\tauBP$.
For example, the noninteracting wave function, which disregards the
two-body interaction, is more extended (in its density profile) than GP.
Its mixed estimator is therefore worse than that with the GP trial wave
function.
The mixed-estimator for the ground-state energy is exact in both, but the
variance is slightly larger with the former.

We now show results for different systems with $N$ from 2 to 9 bosons, and
varying interaction strengths.
We note that if we keep the product $U \times (N - 1)$ constant, the
Gross-Pitaevskii equation predicts the same \emph{per-particle} energies
and densities. For brevity, we shall refer to the curve in which
$U \times (N-1)$ is constant as the \emph{GP isoline}.
Deviation from the GP isoline is therefore an indication of the effect of
many-body correlations.
In order to show results on multiple systems at the same time we will scan
GP isolines.
\begin{figure}[!htbp]

  \includegraphics[scale=0.66]{\PLOTFILE{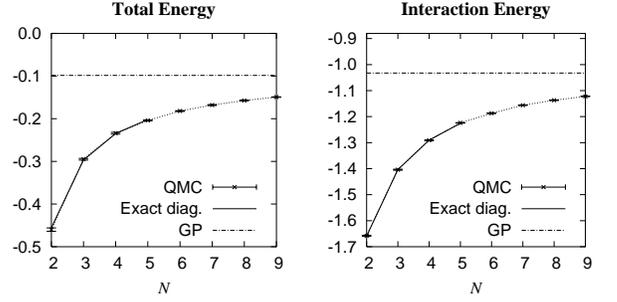}}
  \caption{\label{fig:iso1-1.0-energies}
  Comparison of QMC, GP, and ED results for different systems.
  Calculations were done along a GP isoline $U \times (N - 1) = -2.30 t$
  for up to nine particles in 13 sites. The graphs show the total and
  interaction energies \emph{per particle}.
  QMC and exact results are indistinguishable. GP is accurate in the
  limit of weak correlation but deviates more from the exact results
  as the system becomes more correlated.
  The solid lines are to aid the eye.
  }

%
%
%
\end{figure}
Figure~\ref{fig:iso1-1.0-energies} shows the QMC and GP results as a
function of the number of particles. In the GP calculations the
per-particle quantities are constants.
The QMC results, on the other hand, capture the
effect of correlation. Both the total energy and the interaction
energy are lowered from the GP results. The exact results deviate
from GP more as the system becomes more correlated along the GP
isoline, i.e.~when $U$ is increased or when $N$ is
decreased. Although $N$ is too small here because of the limitation of
ED, the results are representative of the general trend in larger
systems (see below).

Figure~\ref{fig:iso1-1.0-profiles} further illustrates the effect of
particle correlation in this system.
Although the exact interaction energy is lower than that of GP, the exact
density profile is more extended.
This is also manifested in the average trap potential energy
$\trapXV / N$, where the QMC results are $0.1981(8)$ and $0.1605(2)$ for
$N = 2$ and $9$ particles, respectively, while the GP value is $0.1501$.
In GP, interaction energy is lowered by increasing particle overlap,
namely by shrinking the profile. In reality, the particles find a way
to reduce interaction without statically confining to the central sites,
resulting in a more extended one-body profile.

\begin{figure}[!htbp]

  \includegraphics[scale=0.61]{\PLOTFILE{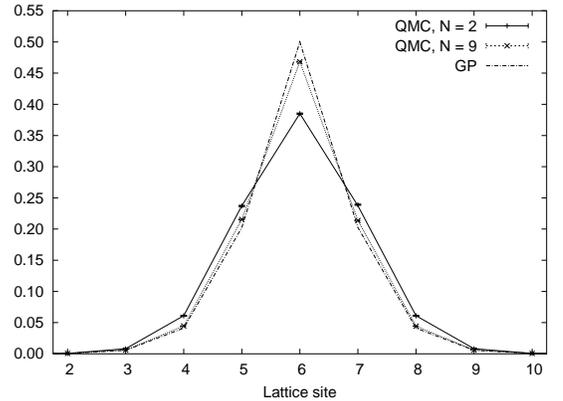}}
  \caption{\label{fig:iso1-1.0-profiles}
  The normalized density profiles as an illustration of
  particle correlation effects.
  Results are for 13-site systems along the GP isoline
  $U \times (N - 1) = -2.30 t$.
  The normalized GP curve is identical for any number of particles
  along this line. QMC results are shown for $N=2$ and $N=9$.
  The QMC results have very small errorbars and are indistinguishable
  from ED (not shown). The QMC density profiles are more extended,
  although the interaction energies are lower than GP, as shown in
  Fig.~\ref{fig:iso1-1.0-energies}.
  }

\end{figure}

\subsection{Comparison with analytic results in 1-D: $a_s < 0$}

The problem of an arbitrary number of untrapped bosons interacting
with an attractive $\delta$-potential in one dimension can be solved
analytically\Cite{McGuire64}, yielding
analytic expressions for the total energy and density profile.
In this section we carry out QMC and GP calculations and compare our
results against these analytic results, on systems of up to 400 bosons.
The Hamiltonian in the continuous real space is%
\EXTRA{\footnote{
    Unlike in the original papers,
    we multiply the Hamiltonian by an extra factor of $\half$ so that its
    kinetic term has a prefactor $\half$, just like our Hamiltonian. See,
    for example, the definition of $t$ in our Hubbard model,
    Eq.~\eqref{eq:Hub2real-t}).
    Because of this, the expression of the energy in
    Eq.~\eqref{eq:Untrapped-g.s.energy} also carry the same prefactor.
}}{}
\eql{eq:Untrapped-H}
{
    \Hop = -\Half \sum_{i = 1}^{N} \frac{\partial^2}{\partial x_i^2}
           -\Half\, g \! \sum_{i>j = 1}^{N} \delta(x_i - x_j) \,.
}
The interaction constant ($g > 0$) is related to our Hubbard parameters by
$g \equiv |U / \sqrt{t^{\,}}|$.
The ground state of this Hamiltonian is an $N$-boson bound state.
\EXTRA{
The condensate, however, can freely ``slide'' like a droplet in the
absence of the external confining potential.
We therefore need to subtract the center-of-mass motion, or equivalently,
fix the center of mass at $x = 0$.
}{}
By fixing the center of mass at $x = 0$, we can eliminate the contribution
from its overall\EXTRA{\ (center-of-mass)}{} motion,
which leads to the following
analytic expressions for the density profile\Cite{Calogero75},
\eql{eq:Untrapped-density}
{
    \rho(x) = \half g \sum_{n=1}^{N-1} (-1)^{n+1}
              \frac{ n (N!)^2 e^{-gnN|x|/2} }
                   { (N+n-1)! (N-n-1)! } \,,
}
and the total energy,
\eql{eq:Untrapped-g.s.energy}
{
    E = - {\textstyle\frac{1}{96}} g^2 N(N^2 - 1) \,.
}

In our QMC calculations, we again put the system on a real-space lattice.
The lattice size is chosen to be large enough so that discretization
errors are comparable to or smaller than statistical errors.
As the ground state of the system is a droplet in the absence of the
external confining potential, the center of mass can slide in the
calculation due to random noise.
We therefore need to subtract the center-of-mass motion.
Technically, this can be accomplished conveniently in the random walk
by treating the system with respect to its center of mass.
In Appendix \ref{app:dplt-crx}, we describe our method for this
correction, which is applicable in any situation where the center of mass
and relative motions need to be separated.
In our calculations, the correction affects the kinetic and total energies
as well as the density profiles. The results shown below were all obtained
with such a correction applied.

\begin{figure}[!htbp]

  \includegraphics[scale=0.685]{\PLOTFILE{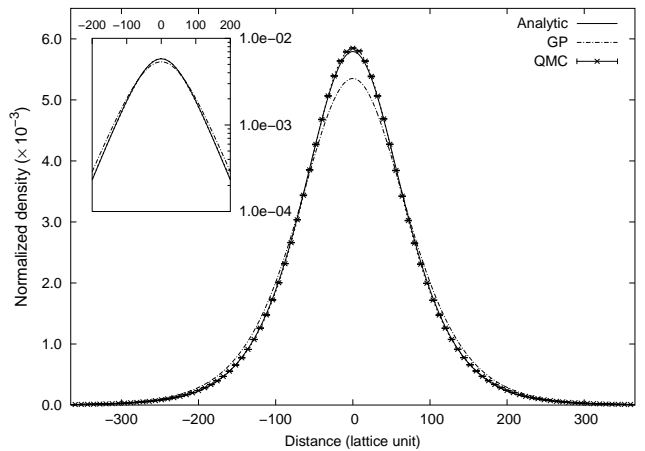}}
  \caption{\label{fig:T1D-1024s20p.iso0.55-profile}
  Comparison of calculated density profiles from QMC and GP
  with analytical results. The densities are normalized.
  The QMC errorbars are displayed every five data points to avoid
  cluttering the plot.
  The QMC profile is given by the \PLOTCOLOR{blue curve}{dotted curve}.
  The inset shows the same curves with logarithmic vertical scale,
  indicating that at large distances the density is exponential.
  }

\end{figure}

We first study a system of 20 particles with $g = 0.154$.
Table~\ref{tbl:T1D-1024s20p.iso0.55-energies} shows the energies, and
\Fig{fig:T1D-1024s20p.iso0.55-profile} the density profiles.
This is a system where mean-field makes significant errors.
Our QMC results are in excellent agreement with the exact results.

\begin{table}[!htbp]
\begin{center}

  \caption{\label{tbl:T1D-1024s20p.iso0.55-energies}
  Comparison of QMC and GP results to available exact results.
  The system has 20 particles and $g = 0.154$. A lattice of 1024
  sites was used, with $\Dt = 0.01$ and $\tauBP = 2.5$.
  }

  \begin{tabular}{llllll}
  \hline
   Type     &  g.s.energy   &  $\kinXV$    &  $\potXV$     & cond.frac. \\
  \hline
  \hline
   Analytic result
            & $-1.971$      &  -           &  -            & -          \\
   QMC      & $-1.964(8)$   & $2.044(8)$   & $-4.007(4)$   & $99.76\%$  \\
   GP       & $-1.784$      & $1.776$      & $-3.561$      & $100\%$    \\
  \hline
  \end{tabular}
\end{center}
\end{table}

\begin{figure}[!htbp]

  \includegraphics[scale=0.7]{\PLOTFILE{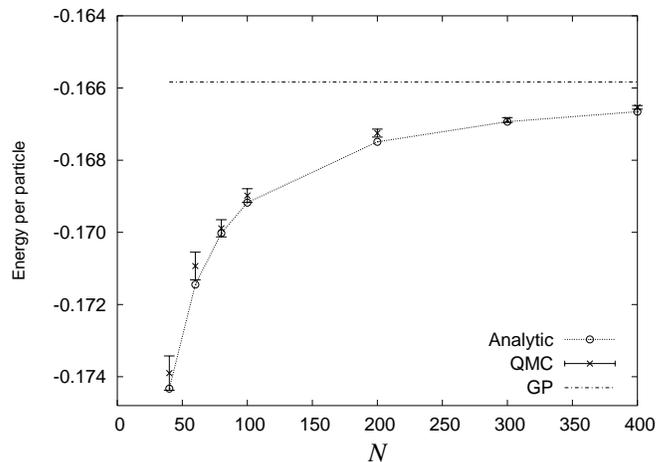}}
  \caption{\label{fig:T1D-1024s.iso0.75-energies}
  Comparison of the energy from QMC (\PLOTCOLOR{blue points}{crosses})
  with the exact answer (\PLOTCOLOR{red}{dotted} curve)
  for different number of particles.
  Energy per particle is shown along the GP isoline
  $g \times (N-1) = 4.0$.
  The GP result is the flat, \PLOTCOLOR{pink}{dash-dotted} line.
  We use a lattice of 1024 sites, $\Dt = 0.01$ and $\tauBP = 4.0$.
  }

\end{figure}

\begin{figure}[!htbp]

  \includegraphics[scale=0.67]{\PLOTFILE{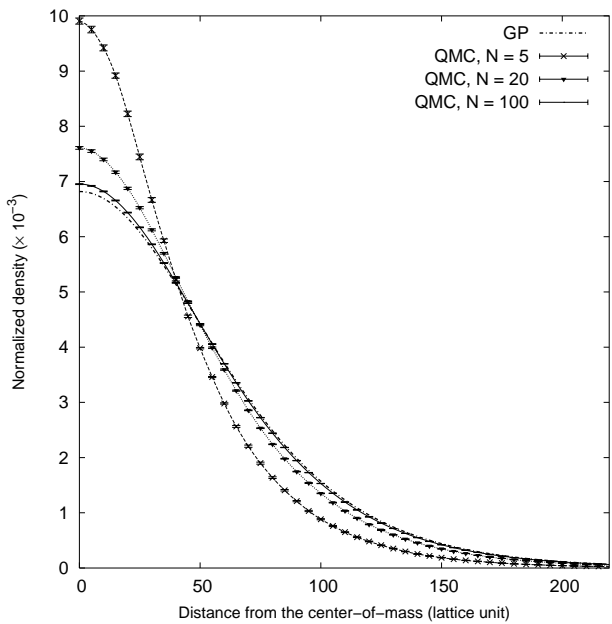}}
  \caption{\label{fig:T1D-1024s.iso0.75-profiles}
  Comparison of the density profiles from QMC and GP with analytic results.
  The normalized densities are shown along the GP isoline
  $g \times (N-1) = 4.0$ for several $N$ values. The system is the
  same as that in \Fig{fig:T1D-1024s.iso0.75-energies}.
  The GP density is the same for any $N$ on the
  isoline, and is given by the \PLOTCOLOR{green}{dash-dotted} line.
  }

\end{figure}

We next scan systems with various numbers of particles by following
the GP isoline $g \times (N - 1) = 4.0$.
The energy per particle is shown as a function of $N$ in
\Fig{fig:T1D-1024s.iso0.75-energies}, for up to 400 particles.
\Fig{fig:T1D-1024s.iso0.75-profiles} shows the density profiles for up
to 100 particles. Again, the agreement between QMC and exact results is
excellent. As the interaction strength $g$ is increased or as $N$
is decreased, mean-field results deviate more and more from the exact results.
For example, as we go from $g=0.01$ ($N = 400$) to 10 times the strength
along the isoline, the systematic error in the GP total energy increases
roughly from $0.5\%$ to $5\%$.

%

\begin{figure}[!htbp]

  \includegraphics[scale=0.7]{\PLOTFILE{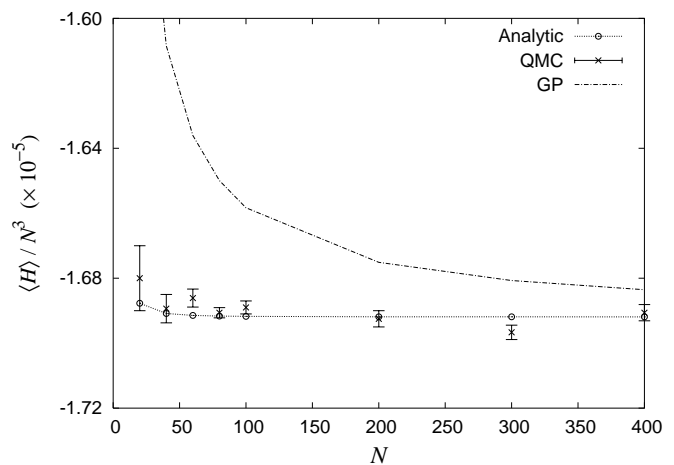}}
  \caption{\label{fig:T1D-1024s.Ux0.0303-energies}
  Comparison of computed ground-state energy for different numbers of
  particles $N$. The interaction strength is held constant at
  $g = -0.0403$.
  The total energy divided by $N^3$ is shown as a function of $N$ for
  QMC, GP and exact calculations.
  Conservative parameters were used, with $\tauBP = 4.0$ in all case,
  and $\Dt = 0.01$ for $N < 200$ and $\Dt = 0.005$ otherwise.
  }

\end{figure}

We now study the system along a different line, holding the interaction
strength $g$ fixed while scanning the number of particles, again up to
$N=400$ particles.
Figure~\ref{fig:T1D-1024s.Ux0.0303-energies} shows the behavior of
$\XV{\Hop} / N^3$ for up to 400 particles, with $g = 0.0403$.
At large $N$, the total energy is roughly proportional to $N^3$.
Compared to Figs.~\ref{fig:T1D-1024s.iso0.75-energies} and
\ref{fig:T1D-1024s.iso0.75-profiles}, the interaction strength here is
stronger at larger $N$ and weaker at lower $N$, with the crossover at
$N\sim 100$.
Most of the calculations are therefore more challenging numerically.
Again QMC was able to completely recover the correlation energy
missed by GP.
At large $N$, smaller timesteps were used and more computing
was necessary to reduce the statistical errors.
(Note that the errorbars appear larger at smaller $N$ in the plot because
of the division by $N^3$.)

\subsection{Comparison with exact diagonalization: $a_s>0$}
\label{ssec:Initial-repulsive}

We have shown that our new QMC algorithm is exact and works well for a
wide range of systems with attractive interactions.
If the interaction is repulsive ($a_s > 0$, or equivalently $U > 0$)
the one-body propagators resulting from the HS transformation become
\emph{complex}, in the form of $\exp({i \sqrt{\Dt U} x_i \nop{i}})$.
The same algorithm applies in this case as well.
In principle the complex one-body operator only requires a change to the
corresponding complex operations.
But in practice a serious phase problem occurs, which causes the
calculation to lose efficiency rapidly at larger interaction strengths.
We discuss this problem and how to control it below.
Our initial studies indicate that, for moderate interaction strengths,
the algorithm as is remains very efficient and gives accurate results,
allowing reliable calculations for parameters corresponding to
experimental situations in 3-D.

We benchmark our algorithm in one- and two-dimensional systems with
repulsive interactions against exact diagonalization.
Table~\ref{tbl:T1D+13s4p.Ux1.0}
shows results for a one-dimensional system, with 13 sites and 4 particles.
The agreement between QMC and exact result is excellent.
Results from GP are also shown.
The GP and QMC density profiles have roughly the same size, as evident
from the values of $\trapXV$.
However, GP overestimates the interaction energy because it does not take
into account the particle-particle correlation.
In the mean field picture, expanding the density profile is the only way
to lower the interaction energy, so that the particles overlap less with
each other. (Note that $\trapXV$ is indeed slightly larger for GP.)
In reality, particles can avoid each other more effectively by means of
many-body correlation. The QMC correctly recovers this correlation, which
lowers the total energy without spreading the density as much as GP does.

\begin{table}[!htbp]
\begin{center}

  \caption{\label{tbl:T1D+13s4p.Ux1.0}
  Comparison of QMC results against exact diagonalization (ED) and
  Gross-Pitaveskii (GP) in 1-D.
  Here we use 13 sites and 4 particles; $t = 2.676$, $U = +1.538$,
  $\kappa = 0.3503$; $\Dt = 0.01$ and $\tauBP = 2.5$.
}

  \begin{tabular}{llllll}
  \hline
   Type     &  g.s.energy   &  $\kinXV$    &  $\trapXV$   &  $\potXV$     & cond.frac. \\
  \hline
  \hline
   ED       &  $4.24$       &  $1.18$      & $1.793$      &  $1.269$      & $98.5\%$  \\
   QMC      &  $4.24(2)$    &  $1.18(2)$   & $1.790(8)$   &  $1.273(8)$   & $98.6\%$  \\
   GP       &  $4.43$       &  $1.03$      & $1.800$      &  $1.599$      & $100\%$    \\
  \hline
  \end{tabular}
\end{center}
\end{table}

Table \ref{tbl:T2D+4s4p.Ux1.0} shows results for bosons in a
two-dimensional trap, using a $4 \times 4$ lattice.
The GP solution also exhibits the same behavior as in the 1-D calculation,
in that the density profile is slightly more extended, and the interaction
energy is overestimated.
As in other cases, the QMC statistical errorbar on the condensate fraction
was not computed directly, but we estimate it to be on the last digit.
\COMMENTED{
Note that the effective interaction strength is not only determined by
the $U/t$ ratio, since the dimensionality of the problem also plays a
role. Higher dimensionality has more degrees of freedom for the kinetic
energy, which results in a weaker effective interaction strength.
}

\begin{table}[!htbp]
\begin{center}
  \caption{\label{tbl:T2D+4s4p.Ux1.0}
  Comparison of QMC calculations against exact diagonalization (ED) and
  Gross-Pitaveskii (GP) projection in a $4 \times 4$ lattice, with 4
  bosons. $t = 0.2534$, $U = +0.3184$, $\kappa = 3.700$; $\Dt = 0.01$ and
  $\tauBP = 2.5$.
  }

  \begin{tabular}{llllll}
  \hline
   Type     &  g.s.energy   &  $\kinXV$    &  $\trapXV$   &  $\potXV$     & cond.frac. \\
  \hline
  \hline
   ED       &  $6.000$      &  $1.818$     & $3.8326$     &  $0.350$      & $97.8\%$  \\
   QMC      &  $6.005(6)$   &  $1.817(2)$  & $3.8325(2)$  &  $0.355(5)$   & $97.8\%$  \\
   GP       &  $6.067$      &  $1.763$     & $3.8359$     &  $0.469$      & $100\%$    \\
  \hline
  \end{tabular}
\end{center}
\end{table}


As mentioned earlier, the only modification necessary to the algorithm in
order to treat repulsive interactions ($a_s > 0$) is to allow complex
arithmetic.
A more serious problem can occur, however.
The orbitals and the walker weights become complex numbers.
Asymptotically the phase of these weights will be uniformly distributed in
the complex plane.
The denomitors in \Eqs{eq:mixed-est-imp} and \eqref{eq:bp-est} will be
dominated by noise, causing the Monte Carlo sampling efficiency to decay
and ultimately destroying the algebraic scaling of QMC.
This is the so-called sign or phase problem\Cite{Zhang1997_CPMC,Zhang2003}.
In real-space methods this problem is connected to fermions, but here we
have a situation where a phase problem appears in the ground state of
a bosonic system. 
Physically, it is easy to see why a phase problem must occur.
Our many-body wave function is being represented in IOR, with only
one orbital in each walker.
With a repulsive interaction, the only way to reflect correlation effects,
i.e., particles avoiding each other, is to make the orbitals complex.

As we see below, our algorithm remains efficient and gives accurate
results for large systems with scattering lengths corresponding to
experimental situations in 3-D.
As the interaction strengths become much stronger, the phase problem will
ultimately make the approach ineffective. We have done preliminary
calculations in which we control the phase problem by
applying a phaseless formalism described in \Ref{Zhang2003}.
Our results indicate that the systematic errors introduced by the
phaseless approximation are small for moderate interaction strengths.
We expect to therefore be able to obtain accurate and reliable results for
scattering lengths well into the experimental 'strong-interaction' regime
achievable by Feshbach resonnance.

\subsection{Realistic calculations in three-dimensions}
\label{ssec:Initial-3D-calc}

\EXTRA{The most exciting problem to address with this model is the study
of Bose-Einstein condensate model in three-dimensional trap.}{}
In this section we present some test results on realistic systems of
trapped particles in three-dimensions.
QMC results were obtained with back-propagation and conservative choices
of $\Dt$ and convergence parameters. We expect the QMC results to be exact.
We also carry out the corresponding Gross-Pitaevskii calculations, and
make comparisons against our exact QMC results.

\begin{figure}[htbp!]

  \includegraphics[scale=0.6]{\PLOTFILE{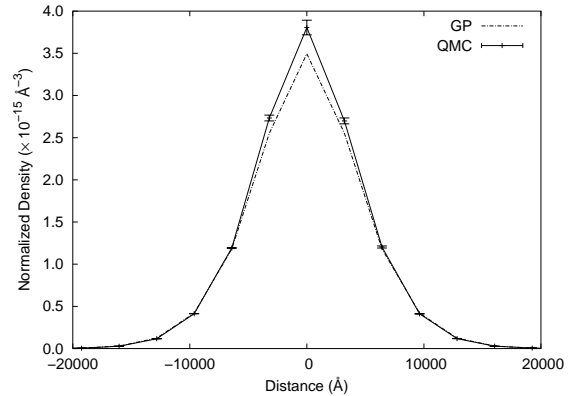}}
  \caption{\label{fig:T3D-15s175p.iso0.375-profiles}
  Comparison of density profiles from the QMC and GP for 175 particles.
  The system is the same as described in
  Table~\ref{tbl:Rb85-15s175p.iso0.375}.
  The QMC profile is more peaked and tighter than GP.
  }

\end{figure}

Table~\ref{tbl:Rb85-15s175p.iso0.375} shows the result of a QMC
calculation for 175 particles in a three-dimensional trap.
We choose a trap with a characteristic length $\aho = 8546$~\AA{}.
The trap was discretized into a $15 \times 15 \times 15$ lattice,
in a range that corresponds to about $5 \times \aho$.
The scattering length is $\as = -22.4$~\AA{}.
In this regime the GP solution is a good approximation to the exact
ground-state wave function. We see that this is indeed the case in
Table~\ref{tbl:Rb85-15s175p.iso0.375}.
The interaction energy is lowered in the many-body calculation as
expected.
Interestingly, the external potential energy is lower than in GP.
Consistent with this, the exact density profile is tighter than
in GP, as shown in \Fig{fig:T3D-15s175p.iso0.375-profiles}.
The trend here appears different from what we observed in small
1-D trapped systems in \Fig{fig:iso1-1.0-profiles}, but consistent with
the large untrapped systems in \Fig{fig:T1D-1024s.iso0.75-profiles}.
We are presently carrying out more calculations to cover a wider range of
parameters and study the role of dimensionality.

\begin{table}[!htbp]
\begin{center}

  \caption{\label{tbl:Rb85-15s175p.iso0.375}
  Comparisons of QMC and GP calculations for 175 particles in a
  3-D spherical trap, with
  $\as = -22.4$~\AA{} and $\aho = 8546$~\AA{}.
  The energies are displayed as per-particle quantities.
  Both the QMC and GP results are extrapolated to $\Dt \rightarrow 0$.
}

  \begin{tabular}{llllll}
  \hline
   Type     &  g.s.energy   &  $\kinXV$    &  $\trapXV$   &  $\potXV$     & cond.frac. \\
  \hline
  \hline
   QMC      &  $16.979(6)$  &  $16.47(5)$  &  $6.54(1)$   &  $-6.03(4)$   & $99.73\%$ \\
   GP       &  $17.115$     &  $15.60$     &  $6.77$      &  $-5.25$      & $100\%$ \\
  \hline
  \end{tabular}
%
%
\end{center}
\end{table}

We now turn to bosons with repulsive interactions in three-dimensional trap.
We again use a $15 \times 15 \times 15$ lattice, and simulate $100$ bosons.
We choose a scattering length $\as$ of $80$~\AA{}.
This value is close to the experimental
\textsuperscript{39}K singlet\Cite{Bohn1999} or
\textsuperscript{87}Rb triplet\Cite{Weiner1999}
scattering lengths.
In Table~\ref{tbl:T3D+15s100p.as80} we show the calculated energies
and condensate fraction.
For this interaction strength, the impact of the phase problem on the
statistical error is small, and the QMC calculation is very efficient.
The true condensate is, like in the 1-D repulsive case, tighter than that
predicted by GP, with lower interaction energy.

\begin{table}[!htbp]
\begin{center}

  \caption{\label{tbl:T3D+15s100p.as80}
  QMC calculation of 100 particles in a three-dimensional trap. A
  lattice of $15 \times 15 \times 15$ was used.
  The parameters correspond to $\aho = 8546$~\AA{} and $\as = 80$~\AA{}.
  The quantities displayed are for per particle.
  }

  \begin{tabular}{llllll}
  \hline
   Type     &  g.s.energy   &  $\kinXV$    &  $\trapXV$   &  $\potXV$     & cond.frac. \\
  \hline
  \hline
   QMC      &   $24.687(9)$ &  $9.573(9)$  &  $11.933(5)$ &  $3.181(3)$   & $99.80\%$  \\
   GP       &   $24.922$    &  $9.281$     &  $12.028$    &  $3.612$      & $100\%$    \\
  \hline
  \end{tabular}

\end{center}
\end{table}

\section{Discussions}\label{sec:Discussions} 

\subsection{Connection between QMC and Gross-Pitaevskii projections}

The QMC method we have presented allows us to go beyond mean-field and
treat many-body effects.
On the other hand, it has a deep connection with the GP mean-field approach.
Our approach uses an HS transformation which leads to integrals of
single-particle operators over auxiliary-fields.
The mean-field solution can be regarded as the leading term in
the stationary-phase asymptotic expansion of the exact
solution\Cite{Negele1998}. Our method evaluates this exact solution,
which is in the form of many-dimensional integrals, by Monte Carlo.
In this section we further comment on the formal connection between our
importance-sampled QMC and the GP as done by projection (the second of
the two GP methods
discussed in subsection~\ref{ssec:GP-implement}).

Let us reconsider the two-body propagator in the modified AF
transformation \Eq{eq:HSxformn-alt}.
Let us suppose that we are now taking our first Monte Carlo step, where
our walker is $\ket{\phi}$, and we will also use the same wave function
as $\ket{\PsiT}$.
Following the discussion of the optimal choice of $\Vxbar$ in the same
section, \ref{ssec:ImpSamp}, we know that $\vec{x} = \mathbf{0}$ is a
stationary point with the choice
\eql{eq:MF-vbar}
{
    \xbar[i]
  = -\sqrt{\Dt} \, \vbar[i]
    \equiv
    -\sqrt{\Dt} \, \frac{ \ME{\phi}{\hat{v}_i}{\phi} } 
                        { \braket{\phi}{\phi} }
    \,.
}
We can approximate the integral in \Eq{eq:HSxformn-alt} by the value of
the integrand at $\vec{x} = \mathbf{0}$, which can be justified in the
limit of small $\Dt$.
More explicitly, as $\Dt \rightarrow 0$, the Gaussian function becomes the
most rapidly varying term in the integrand.
To exhibit the asymptotic behavior of this integral, we change
the integration variable to
$
    \vec{y} \equiv \sqrt{\Dt} \, \vec{x}
$,
so that the large parameter $1 / \Dt$ appears in the Gaussian's exponent:
\eq{
    \expP{\half\Dt \hat{v}^2}
&   =
    e^{-\Dt(\Half\vbar[]^2 - \vbar[]\hat{v})}
    \int_{-\infty}^{\infty} dy \,
    \frac{e^{-y^2 / 2\Dt}}{\sqrt{2\pi\Dt}}
    e^{y(\hat{v}-\vbar[])}
    \,.
}
The dominant contribution to the integral comes from the maximum of the
Gaussian function at $y = 0$. The asymptotic leading term of the
importance-sampled many-body propagator is therefore:
\eql{eq:MF-asymp-Pgs}
{
    \expB{-\Dt \big(\Kop
                  - \sum_i \vbar[i] \hat{v}_i
                  + \Half \sum_i \vbar[i]^2
               \big)}
    \,,
}
where $\Kop$ is the one-body term in the original Hamiltonian.
Under this approximation, our random walk becomes deterministic, needing
only one walker.
If for the next step we use the updated wave function $\ket{\phi'}$ to
evaluate the new $\{\vbar[i]\}$ in \Eq{eq:MF-vbar}, we obtain a
self-consistent projection with one-body propagators.
In fact, the one-body Hamiltonian in the exponent of \Eq{eq:MF-asymp-Pgs}
is precisely the mean-field Hamiltonian.
For example, for Bose-Hubbard model the last two terms in the exponent
lead to the GP mean-field potential
\eql{eq:MF-H2B-asymp}
{
    U \sum_i \Big(\nbar_i \nop{i} - \half \nbar_i^2 \Big) \,.
}
Apart from the factor $(N-1)/N$ which approaches unity in the limit of
large $N$, we have recovered the GP propagator.
The projection with \Eq{eq:MF-asymp-Pgs} lowers the variational energy for
any initial $\ket{\phi}$ and is stationary when $\ket{\phi}$ is the GP
solution.
This is why GP is the best variational wave function that has the form of
a single permanent, and hence a reasonable trial wave function to use for
most of our QMC calculations.

It is also clear from the discussion above that the importance sampling
formalism allows us to have an optimal form of HS transformation, in that
the HS propagator $e^{y(\hat{v} - \vbar[])}$ involves only the difference
$\hat{v}-\vbar[]$.
In other words, although in \Eq{eq:HSxformn-general} we write the
decomposition for the bare interaction term, the importance sampling
transformation effectively introduces a mean-field background based on
the trial wave function and allows the HS to deal with only a residual
quadratic interaction term, $(\hat{v}-\vbar[])^2$.

To summarize, our QMC method reduces to GP if we evaluate the many-body
propagator by the stationary-point approximation, using only the centroid
of the Gaussian.
The full method evaluates the many-dimensional integral over
auxiliary-fields exactly by Monte Carlo.
It captures the interaction and correlation effects with a stochastic,
coherent ensemble of mean-field solutions.
The structure of the calculation can be viewed as a superposition of the
GP projections that we have described.
Our method therefore provides a way to systematically improve upon GP
while using the same framework.

\subsection{Computing}

Because of the structure of QMC as a superposition of GP projections, our
method scales gracefully with system size.
As discussed in \Sec{ssec:QMC-implement}, the bulk of our method scales as
$\Order(M \log M)$, with the significant speedup from using Fast Fourier
transform.
For example, the QMC calculation shown in Table~\ref{tbl:T3D+15s100p.as80}
required less than 8 hours on a single Alpha EV67 processor.
The 1024-sites QMC calculation shown in
Table~\ref{tbl:T1D-1024s20p.iso0.55-energies} took about four hours
to get good statistics, with very conservative choices of $\Dt$ and other
convergence parameters. It required about 1.3 gigabytes of memory, largely
because of back-propagation path recording.
In contrast, treated fully, the latter problem would mean the
diagonalization of a sparse, Hermitian matrix containing
$(8 \times 10^{41})^2$ elements.
Although this can be reduced by exploiting symmetries, exact
diagonalization of this problem is clearly not within reach with
computing capabilities in the foreseeable future.
%
%
%
%
%
%
%

We typically use hundreds of walkers in our calculation.
The stochastic nature of QMC means the number of walkers fluctuates due to
branching and killing of walkers with very large and very small weights
(see subsection~\ref{ssec:BRW}).
The population therefore must be controlled to ensure that it does not
grow or decay too much, and that the walker weights have a reasonable
distribution. Our method to control the population is similar to that
discussed in \Ref{Zhang2000_BookChapter}.

We comment on the effect of the number of particles, $N$, on scaling.
Because of the use of IOR, the number of particles does not enter in the
propagation. It would then seem as though the algorithm might have a
super-scaling in $N$.
This is not true, of course, since the projector $\expdtH$ depends on $N$.
For example, the shift $\vbar[i]$ has a factor of $N$ in front (see
Appendix~\ref{app:IOR}), and the local energy scales with $N$.
As a result, a smaller time-step must be used for larger $N$.
The above arguement suggests a linear reduction in $\Dt$ as $N$ is
increased, which we have used as a rough guideline in our calculations
to select the range of $\Dt$ to use.
Extrapolations with separate calculations using different $\Dt$ values are
then carried out.

\subsection{Conclusion and Outlook}\label{ssec:Conclusion}

In conclusion, we have presented a new auxiliary-field QMC
algorithm for obtaining the many-body ground state of bosonic systems.
The method, which is based upon the field-theoretical framework and
is essentially exact, provides a means to treat interactions more
accurately in many-body systems.
Our method shares the same framework with the GP approach, but captures
interaction and correlation effects with a stochastic ensemble of
mean-field solutions.
We have illustrated our method in trapped and untrapped boson atomic gases
in 1-, 2-, and 3-dimensions, using a real-space grid as single-particle
basis which leads to a Bose-Hubbard model for these systems.
We have demonstrated its ability to obtain exact ground-state properties.
We have also carried out the GP mean-field calculations and compared the
predictions with our exact QMC results.
Our method is capable of handling large systems, thus providing the
possibility to simulate system sizes relevant to experimental situations.
We expect the method to complement GP and other approaches, and
become a useful numerical and theoretical tool for studying trapped atomic
bosons, especially with the growing ability to tune the interaction
strengths experimentally and reach more strongly interacting regimes.


From the methodological point of view, more work remains to be done
with the repulsive case to deal with the phase problem. We have shown
that our method as it stands can be very useful for moderate interaction
strengths.
For stronger interactions, our preliminary study indicates that the
phaseless approximation\Cite{Zhang2003}, which eliminates the phase
problem but introduces a systematic error, is very accurate for scattering
lengths well into the Feshbach resonnance regime.
We are currently examining this more systematically to quantify the extent of
the bias.
Because of the simplicity of these bosonic systems compared to electronic
systems, they provide an ideal testbed, where for small sizes the problem
is readily solved by exact diagonalization.

A variety of applications are possible.
The ground state of the Bose-Einstein condensates with both attractive and
repulsive interatomic interactions can be studied for various interaction
strengths, including the strongly interacting regime reached by Fesbach
resonance.
They can also be studied in different dimensions and under different
conditions.
In particular, it would seem straightforward to generalize our present
framework to study rotations and vortices, since we are already dealing
with complex propagators and wave functions in the repulsive case.
In addition, it will be interesting to treat boson-fermion mixtures with
our approach. As mentioned, the auxiliary-field method is already widely
used to treat strongly interacting fermion systems.


\begin{acknowledgments} 

We thank D.~M.~Ceperley and H.~Krakauer for stimulating discussions.
Financial support from NSF (grant DMR-9734041), ONR (grant
N00014-97-1-0049),
and the Research Corporation is gratefully acknowledged.
SZ expresses his gratitude to Prof.'s Ceperley and Richard Martin for their
hospitality during a sabbatical visit, where part of the work was carried
out.
We also thank the Center of Piezoelectric by Design (CPD), where part of
our computing was performed.

\end{acknowledgments}

\appendix 

\section{Identical-orbital representation}\label{app:IOR}

In this appendix we show that the matrix representation of an
$N$-boson wave function in AFQMC can be made particularly simple.
In fermion calculations, we must use an $M\times N$ matrix to represent
a determinant, because the orbitals must be mutually orthogonal.
In the boson case, however, this restriction is absent.
The most general form of a many boson permanent is expensive to compute,
having complexity of $\Order(NM!)$.
But we can choose to make all the orbitals identical.
In matrix language, we will have only an $M$-row column vector.
We will term this representation
\emph{identical-orbital representation}---IOR.
Each many-boson wave function in IOR has the form of a
GP mean-field solution.
Two conditions are necessary for this choice to be viable in the QMC:
that an initial trial wave function of this form is allowed and that
successive projections preserve the form.
The only requirement for the former to hold is that the wave function in
IOR not be orthogonal to the true many-body ground state, and it is
straightforward to show that \Eq{eq:exp-operator} holds for a
$\ket{\phi}$ in this form.
More complex wave functions can always be generated by a linear
combination of such wave functions. In fact, this is what we accomplish
through our Monte Carlo simulation.

In operator language, a single $N$-boson wave function $\ket{\phi}$ is
given by
\eq{
    \ket{\phi} = \underbrace{\Cphi{}\Cphi{}\cdots\Cphi{}}_{N}\ket{0}
               = \big(\Cphi{}\big)^N \ket{0} \,,
}
where $\Cphi{} \equiv \sum_\alpha \Cc{\alpha}\phi\_{\alpha}$. In matrix
form, $\ket{\phi}$ would be $M\times N$ matrix $\matrixg{\phi}$ whose
columns are identical.
The overlap of two such wave functions is given by
\eq{
    \braket{\psi}{\phi} &= \per{\matrixgT{\psi}\cdot\matrixg{\phi}} \\
                        &= N! (\VpsiT{}\cdot\Vphi{})^N\,,
}
where the bold-phased symbols $\Vpsi{}$ and $\Vphi{}$ represent the
single-column vectors for $\psi$ and $\phi$, respectively.
Similarly, for any one-body operator $\hat{A}$,
\eql{eq:IOWF_1body_ME}
{
    \ME{\psi}{\hat{A}}{\phi}
  = N!\,N (\VpsiT{}\cdot\matrixr{A}\cdot\Vphi{})
          (\VpsiT{}\cdot\Vphi{})^{N-1} \,,
}
where $\matrixr{A}$ is the matrix for $\hat{A}$.
The matrix element of a quartic (two-body) operator is given by:
\eql{eq:IOWF_2body_GF_ME}
{
    \ME{\psi}{\Cb{\alpha}\Cb{\beta}\Db{\gamma}\Db{\delta}}{\phi}
& = N!\,N (N-1) \psi_\alpha^* \psi_\beta^* \phi\_\gamma \phi\_\delta
    (\VpsiT{}\cdot\Vphi{})^{N-2} \,.
}

\section{Droplet center-of-mass correction}\label{app:dplt-crx} 

\subsection{Correcting the density broadening}

To handle the droplet system given by the translationally invariant
Hamiltonian in \Eq{eq:Untrapped-H}, an extra ingredient is necessary in
addition to the ``basic'' QMC algorithm that we have described.
In a deterministic calculation, for example in GP, the motion of the
center-of-mass (CM) can be simply eliminated by fixing it at the origin,
as in \Eq{eq:Untrapped-density}.
In the QMC calculation, however, the orbitals fluctuate as they are
propagated by $\Bop(\vec{x} - \Vxbar)$, where the random fields $\vec{x}$
are drawn from a Gaussian probability density.
Random noise will inevitably cause the CM of the system to slide,
undergoing a free diffusion whose average position is the origin.

Left unchecked, this spurious CM motion will lead to an artificial
broadening of the density profile. To correct for it in the density
profile, we could simply shift the CM of every walker back to the origin.
However, the importance-sampled propagator involves ratios of overlaps
with the trial wave function $\braket{\PsiT}{\wlkr{i}{}}$, which would
have to be corrected in the random walk whenever a shift is made.

Instead our solution to this diffusive motion is to let the trial wave
function slide along with the walkers. In other words, we rewrite the
kinetic energy operator as
\eql{eq:kinCM-sep1}
{
    \Top = \Topcm + \Top' \,,
}
where $\Topcm$ represents the CM kinetic energy, and $\Top'$ the internal
kinetic energy \emph{in} the CM frame. The total Hamiltonian is given by
\eql{eq:H-kinCM-sep1}
{
    \Hop = \Topcm + \Top' + \Vop
         \equiv \Topcm + \Hop' \,.
}
The quantities that we wish to compute are governed by the ``internal''
Hamiltonian $\Hop'$. Since $\Vop$ involves only relative coordinates among
the particles, it commutes with $\Topcm$; or more generally,
\eql{eq:H-kinCM-commute}
{
    [ \Topcm, \Hop' ] = 0 \,.
}
In this way, the importance-sampled QMC propagation is determined by $\Hop'$.
The motion of the CM in each walker is a separate free diffusion
which is governed by $\Topcm$.
In the random-walk process, we are now free to correct for the CM motion
by shifting the walkers back to the origin whenever necessary.
For consistency, this correction must be applied both in the normal random
walk and in the back-propagation phase.

%
%
%

\subsection{Separating the center-of-mass kinetic energy}

The moving trial wave function, however, poses a problem for the
calculation of the kinetic energy.
Now the orbitals are free to slide, and the diffusive motion of the
orbital's CM is no longer suppressed in the LAB frame.
When we use the usual $t$-term in the Hamiltonian in
\Eq{eq:Hubbard_H_final} to compute the kinetic energy, we obtain the total
$\kinXV$, in which $\Tcm \equiv \XV{\Topcm}$ and the desired $\XV{\Top'}$
are mixed.
This leads to a spurious increase in the estimate of the kinetic energy
and consequently the total energy.
For example, the \emph{uncorrected} ground-state energy for the system
shown in Table~\ref{tbl:T1D-1024s20p.iso0.55-energies} would be
$-1.887(2)$ with $\kinXV = 2.092(3)$; thus the total energy is
overestimated by 0.08 due to the contribution from $\Tcm$.
Since we know the nature of the CM motion, it is fairly straightforward to
extract $\Tcm$ and explicitly subtract it from the kinetic and total
energy estimates.
Allowing the droplet to freely slide in the calculation is equivalent to
having a spurious ``propagator'' $\expP{-\Dt\Topcm}$, whose effect on the
wave function for the CM is described by the diffusion equation
\eq{
    -\frac{\partial\PsiCM(\mathbf{R},\tau)}{\partial\tau}
  = \Topcm \PsiCM(\mathbf{R},\tau) \,.
}
It is a well known property of such a diffusion process that the averaged
squared distance $\XV{\mathbf{R}^2(\tau)}$ grows linearly
with the (imaginary) time $\tau$:
\eq{
    \XV{\mathbf{R}^2(\tau)} = b \tau \,.
}
We can obtain $b$ by recording the quantity $\XV{\mathbf{R}^2(\tau)}$ for
a period of time in the QMC simulation.
The constant $b$ is linearly proportional to $\Tcm$.
More specifically, the center-of-mass Hubbard hopping parameter $\tcm$
can be extracted from $b$:
\eql{eq:b-to-tcm}
{
    \tcm = b / 2 \,.
}
This gives us the correct kinetic and total energies without the
spurious center-of-mass motion:
\eqssl{eq:kinXV-corrected}
{
    \XV{\Top'} &= \left(1 - {\textstyle\frac{\tcm}{t}}\right) \kinXV \,;
\\
    \XV{\Hop'} &= \XV{\Top'} + \potXV \,.
    \label{eq:g.s.energy-corrected}
}

To conclude, there are two necessary modifications in the QMC
algorithm in order to treat quantum droplets which are not confined:
\begin{enumerate}

\item We let the trial wave function effectively ``follow'' the QMC
orbitals, by defining its CM with that of each QMC orbital.

\item For each orbital, we keep track and accumulate all the applied CM
shifts in order to estimate $\XV{\mathbf{R}^2(\tau)}$. This gives us the
fraction of CM kinetic energy through the constant $\tcm$.

\end{enumerate}
These modifications in the QMC allows us to obtain the correct density
profile and energies of a translationally-invariant Hamiltonian.

\bibliography{AFQMC-bib-entries}    

\end{document}